\documentclass[seceq]{ptptex}
\usepackage{wrapft}
\usepackage{graphicx}

\newcommand{\bra}[1]{\langle {#1} |}     %%
\newcommand{\ket}[1]{| {#1} \rangle}     %%
\newcommand{\wtilde}[1]{\widetilde{#1}} %%

\newcommand{\ovl}[1]{\overline{#1}}
\newcommand{\poon}[1]{\stackrel{\rightharpoonup} {#1}} %%
\def\beq{\begin{eqnarray}}
\def\eeq{\end{eqnarray}}
\def\bsub{\begin{subequations}}
\def\esub{\end{subequations}}
\def\b{\begin{equation}}
\def\bs{\begin{split}}
\def\es{\end{split}}
\def\e{\end{equation}}
\begin{document}

\title{A possible framework of the Lipkin model obeying the $su(n)$-algebra in arbitrary fermion number. II
}
\subtitle{Two subalgebras in the $su(n)$-Lipkin model and an approach to the construction of 
linearly independent basis}

\author{Yasuhiko {\sc Tsue}$^{1,2}$, {Constan\c{c}a {\sc Provid\^encia}}$^{1}$, {Jo\~ao da {\sc Provid\^encia}}$^{1}$ and 
{Masatoshi {\sc Yamamura}}$^{1,3}$
%Thanks{These authors contributed equally to this work}}
}
%%%%%%%%%%% The \name command should be used as \name{Insert author name here}{Insert affiliation number here}
%%%%% Please use \thanks for contributed author details

%%%%%%%%%%% The \affil command should be used as \affil{Insert affiliation number here}{Insert author address here}
\inst{$^{1}${CFisUC, Departamento de F\'{i}sica, Universidade de Coimbra, 3004-516 Coimbra, 
Portugal}\\
$^{2}${Physics Division, Faculty of Science, Kochi University, Kochi 780-8520, Japan}\\
%\affil{2}{Departamento de F\'{i}sica, Universidade de Coimbra, 3004-516 Coimbra, 
%Portugal}\\
$^{3}${Department of Pure and Applied Physics, 
Faculty of Engineering Science, Kansai University, Suita 564-8680, Japan}
}

\abst{
Standing on the results for the minimum weight states obtained in the previous paper (I), an idea how to construct the linearly 
independent basis is proposed for the $su(n)$-Lipkin model. 
This idea starts in setting up $m$ independent $su(2)$-subalgebras in the cases with $n=2m$ and 
$n=2m+1$ $(m=2,\ 3,\ 4,\cdots)$. 
The original representation is re-formed in terms of the spherical tensors for the 
$su(n)$-generators built under the $su(2)$-subalgebras. 
Through this re-formation, the $su(m)$-subalgebra can be found. 
For constructing the linearly independent basis, not only the $su(2)$-algebras but also the $su(m)$-subalgebra 
play a central role. 
Some concrete results in the cases with $n=2,\ 3,\ 4$ and 5 are presented. 
}

%\subjectindex{xxxx, xxxx}

%\parindent0pt

\maketitle

\section{Introduction}

This paper, (II), is a continuation of (I) \cite{1} and we will mainly 
discuss an idea, with the aid of which the linearly independent basis is constructed 
for the $su(n)$-Lipkin model. 
As is well known, the $su(n)$-Lipkin model is a classical model for describing certain phenomenon observing in 
nuclei \cite{2}, \cite{3}. 
It is a kind of the Lie algebraic model. 
Therefore, first task, which is inevitable for investigating this model, is construction of the minimum weight states. 
By the definition of the minimum weight states, this task may be performed independently of any Hamiltonian 
expressed as a function of the generators. 
In (I), we proposed an idea how to construct the minimum weight states for the case with any total 
fermion number in $n$ single-particle levels. 
A certain $su(2)$-algebra, which we called the auxiliary $su(2)$-algebra in (I), plays a central role for the minimum weight states. 
The raising operator in this auxiliary algebra is expressed in the form with 
$n$-th degree for the fermion creation operators and the Cliford numbers.

After determining the minimum weight states, it may be natural to proceed to construct the 
linearly independent basis by operating various operators on the minimum weight states such as 
the raising operator in the $su(2)$-algebra. 
However, the following should be noted: 
Generally, it is impossible for determining the orthogonal sets to prepare sufficient number of the hermitian operators, 
which commute with one another. 
Therefore, in this paper, we will aim at constructing the linearly independent basis, and, then, for example, 
with the aid of the Schmidt method, we are able to obtain the orthogonal sets. 
In order to realize this aim, first, we pay attention to the point that, in the $su(n)$-Lipkin model, we can define $m$ independent 
$su(2)$-subalgebras. 
Here, $m$ is an integer related to $n$ under the relation 
$n=2m$ ($n$:even) and $n=2m+1$ ($n$:odd), where $m=2,\ 3,\cdots$. 
Then, by adding these $su(2)$-algebras, we get the $su(2)$-algebra. 
With the use of this $su(2)$-algebra, we can classify all the generators into the scalars, the vectors and the spinors. 
And, we can prove that the set of the scalars gives us the $su(m)$-algebra. 
Therefore, the remaining vectors and spinors may be treated in the space given under the 
$su(2)$- and the $su(m)$-algebra. 
The above may be nothing but the re-formation of the $su(n)$-Lipkin model.

For a given value of the integer $m$ ($m=2,\ 3,\cdots$), we have two Lipkin models: 
the $su(2m)$- and the $su(2m+1)$-algebra. 
As a general argument, we know that the minimum weight states of the $su(n)$-algebra are specified 
by $(n-1)$ quantum numbers. 
In these numbers, $m$ quantum numbers are related to $m$ $su(2)$-subalgebras and, then, 
the remaining ones are $(m-1)$ for $n=2m$ and $m$ for $n=2m+1$, respectively. 
In these numbers, $(m-1)$ quantum numbers may be related to the minimum weight states 
of the $su(m)$-algebra. 
Therefore, in the case of the $su(2m)$-algebra, the number $(n-1)$ is decomposed into $m+(m-1)$. 
In the case of the $su(2m+1)$-algebra, the number $(n-1)$ is decomposed into $m+(m-1)+1$, 
where one quantum number is excess. 
We will show that this extra number is closely related to the spinor operators, because the spinors appear only in the case with $n=2m+1$. 
However, these features may be almost impossible to show in general case, and, then, we will discuss them through the cases with 
$n=2,\ 3,\ 4$ and 5. 
The minimum weight states of these four cases were presented in (I) under the 
re-formed version. 
The cases with $n=4$ and 5 correspond to $m=2$ and they can be treated quite easily, because of $m=2$, i.e., the $su(2)$-algebra.

After giving $m$ $su(2)$-algebras, in {\S 2}, the spherical tensor operators 
(the scalar, the vectors and the spinors) are defined. 
In {\S 3}, the Casimir operator as a quadratic form for the generators is diagonalized and through this procedure, 
the operators specifying the minimum weight states are introduced. 
Of course, the Hamiltonian given in (I) is rewritten in terms of the spherical tensors. 
In {\S 4}, our scheme for obtaining the linearly independent basis is presented. 
Following our scheme, in {\S 5}, the cases with $n=2,\ 3,\ 4$ and 5 are treated. 
In {\S 6}, as a concluding remark, the connection between the $su(2m)$- and the $su(2m+1)$-Lipkin model is discussed. 
In Appendix A, the $su(m)$-subalgebra in the $su(n)$-Lipkin model 
is presented in detail. 
Including the results given in (I), various formal aspects of the Lipkin model is also stressed in (II).

\setcounter{equation}{0}
\section{The $su(n)$-generators of the Lipkin model in spherical tensor form}

In {\S 6} of Part I, we gave some examples of the re-formed generators of the Lipkin model. 
Following these examples, we will treat the present model under two cases: 
(i) $n$; even and (ii) $n$; odd. 
In the case (i), we introduce the integer $m$ through 
\beq\label{2-1}
n=2m\ ; \ \ n=2,\ 4,\cdots ,\ \ \ {\rm i.e.,} \ \ \ m=1,\ 2, \cdots .
\eeq
We also treat the case (ii) by the integer $m$, which is given by 
\beq\label{2-2}
n=2m+1\ ; \ \ n=3,\ 5,\cdots , \ \ \ {\rm i.e.,}\ \ \ m=1,\ 2,\cdots .
\eeq
In (I), we learned that the case with $m=1$ corresponds the $su(2)$- and the $su(3)$-Lipkin model. 
The former is the $su(2)$-algebra itself and the latter contains one $su(2)$-subalgebra. 
On the other hand, the $su(4)$- and the $su(5)$-Lipkin models corresponding to the case with $m=2$ contain 
two $su(2)$-subalgebras. 
Extending this aspect, we define $m$ independent $su(2)$-subalgebras. 
In the case with $n=2m$, the following is given:
\bsub\label{2-3}
\beq
& &{\wtilde S}_+(m;r)={\wtilde S}_{2m-2r}^{2m-2r+1} \ , \qquad {\wtilde S}_-(m;r)={\wtilde S}_{2m-2r+1}^{2m-2r}\ , 
\nonumber\\
& &{\wtilde S}_0(m;r)=\frac{1}{2}\left({\wtilde S}_{2m-2r+1}^{2m-2r+1}-{\wtilde S}_{2m-2r}^{2m-2r}\right)\qquad {\rm for}\ \ \ 
r=1,\ 2,\cdots ,\ m-1\ , 
\label{2-3a}\\
& &{\wtilde S}_+(m;r)={\wtilde S}^1\ , \qquad {\wtilde S}_-(m;r)={\wtilde S}_1\ , \qquad
{\wtilde S}_0(m;r)=\frac{1}{2}{\wtilde S}_1^1\qquad {\rm for}\ \ \ r=m\ . \ \ \ \ 
\label{2-3b} 
\eeq
\esub
The case with $n=2m+1$ gives us the form 
\beq\label{2-4}
& &{\wtilde S}_+(m;r)={\wtilde S}_{2m-2r+1}^{2m-2r+2} \ , \qquad {\wtilde S}_-(m;r)={\wtilde S}_{2m-2r+2}^{2m-2r+1}\ , 
\nonumber\\
& &{\wtilde S}_0(m;r)=\frac{1}{2}\left({\wtilde S}_{2m-2r+2}^{2m-2r+2}-{\wtilde S}_{2m-2r+1}^{2m-2r+1}\right)\qquad {\rm for}\ \ \ 
r=1,\ 2,\cdots ,\ m-1, \ m\ . 
\eeq
We can see that the cases with $(m=1,\ r=1)$ and $(m=2,\ r=1,\ 2)$ reduce 
to the form treated in (I) for the cases with $n=2,\ 3,\ 4$ and 5. 
Of course, we have the relation 
\beq\label{2-5}
& &[\ {\wtilde S}_+(m;r)\ , \ {\wtilde S}_-(m;r')\ ]=\delta_{rr'}\cdot\left(2{\wtilde S}_0(m;r)\right)\ , 
\nonumber\\
& &[\ {\wtilde S}_0(m;r)\ , \ {\wtilde S}_{\pm}(m;r')\ ]=\delta_{rr'}\cdot\left(\pm{\wtilde S}_{\pm}(m;r)\right)\ . 
\eeq
In terms of total sum of ${\wtilde S}_{\pm,0}(m;r)$ for $r$, we obtain 
the $su(2)$-algebra: 
\beq\label{2-6}
{\wtilde S}_{\pm,0}(m)=\sum_{r=1}^m{\wtilde S}_{\pm,0}(m;r)\ . 
\eeq
In our re-formation, the $su(2)$-algebra (\ref{2-6}) will play a role for classifying all generators in terms of the spherical tensor representation.

Main task of {\S 2} is to express all the generators except the forms (\ref{2-3}) and (\ref{2-4}) in spherical tensor forms. 
% with respect to $({\wtilde S}_{\pm,0}(m))$ shown in the relation (\ref{2-3}). 
For this task, it may be convenient to introduce two integers $r$ and $k$:
\beq
& &{\rm (i)}\ \ \ n\ ;\ {\rm even}\ (n=2m)\ , \quad r=1,\ 2,\cdots ,\ m-1 , \quad 
k=r+1,\ r+2,\cdots ,\ m, 
\label{2-7}\\
& &{\rm (ii)}\ \ n\ ;\ {\rm odd}\ \  (n=2m+1)\ , \ \ r=1,\ 2,\cdots ,\ m-1,\ m, \quad 
k=r+1,\ r+2,\cdots ,\ m. \qquad\ \ 
\label{2-8}
\eeq
In the case (ii), if $r=m$, $k$ becomes meaningless. 
In the case with $n=2m$, ${\wtilde S}^p$ and ${\wtilde S}_q^p$ $(p>q)$ can be re-formed to scalars and vectors: 
\beq
{\rm (\alpha)}& & scalars, \nonumber\\
& &{\wtilde R}^{0,0}(m;r,k)=
\left\{
\begin{array}{ll}
{\displaystyle {\wtilde S}_{2m-2k+1}^{2m-2r+1}+{\wtilde S}_{2m-2k}^{2m-2r}} \ , & (k=r+1,\ r+2, \cdots ,\ m-1) \\
\ & \\
{\displaystyle {\wtilde S}_{1}^{2m-2r+1}+{\wtilde S}^{2m-2r}} \ , & (k=m) 
\end{array}\right.
\label{2-9}\\
{\rm (\beta)}& & vectors, \nonumber\\
& &{\wtilde R}^{1,+1}(m;r,k)=
\left\{
\begin{array}{ll}
{\displaystyle -{\wtilde S}_{2m-2k}^{2m-2r+1}} \ , & (k=r+1,\ r+2, \cdots ,\ m-1) \\
\ & \\
{\displaystyle -{\wtilde S}^{2m-2r+1}} \ , & (k=m) 
\end{array}\right.
\label{2-10}\\
& &{\wtilde R}^{1,0}(m;r,k)=
\left\{
\begin{array}{ll}
{\displaystyle \frac{1}{\sqrt{2}}\left({\wtilde S}_{2m-2k+1}^{2m-2r+1}-{\wtilde S}_{2m-2k}^{2m-2r}\right)} \ , & (k=r+1,\ r+2, \cdots ,\ m-1) \\
\ & \\
{\displaystyle \frac{1}{\sqrt{2}}\left({\wtilde S}_{1}^{2m-2r+1}-{\wtilde S}^{2m-2r}\right)} \ , & (k=m) 
\end{array}\right.
\label{2-11}\\
& &{\wtilde R}^{1,-1}(m;r,k)=
\left\{
\begin{array}{ll}
{\displaystyle {\wtilde S}_{2m-2k+1}^{2m-2r}} \ , & (k=r+1,\ r+2, \cdots ,\ m-1) \\
\ & \\
{\displaystyle {\wtilde S}_1^{2m-2r}} \ . & (k=m) 
\end{array}\right.
\label{2-12}
\eeq
In the case with $n=2m+1$, spinors appear:
\beq
{\rm (\gamma)}& & scalars, \nonumber\\
& &{\wtilde R}^{0,0}(m;r,k)=
{\wtilde S}_{2m-2k+2}^{2m-2r+2}+{\wtilde S}_{2m-2k+1}^{2m-2r+1} \ , 
\quad 
\left(
\begin{array}{l}
r=1,2,\cdots, m-1, \\
k=r+1, r+2,\cdots, m
\end{array}
\right)
\label{2-13}\\
{\rm (\delta)}& & spinors, \nonumber\\
& &{\wtilde R}^{\frac{1}{2},+\frac{1}{2}}(m;r)=
{\wtilde S}^{2m-2r+2}\ , \qquad {\wtilde R}^{\frac{1}{2}, -\frac{1}{2}}(m;r)={\wtilde S}^{2m-2r+1}\ , 
\ \ 
(r=1,2,\cdots ,m)\quad
\qquad
\label{2-14}\\
{\rm (\epsilon)}& & vectors, \nonumber\\
& &{\wtilde R}^{1,+1}(m;r,k)=
-{\wtilde S}_{2m-2k+1}^{2m-2r+2} \ , 
\label{2-15}\\
& &{\wtilde R}^{1,0}(m;r,k)=
\frac{1}{\sqrt{2}}\left({\wtilde S}_{2m-2k+2}^{2m-2r+2}-{\wtilde S}_{2m-2k+1}^{2m-2r+1}\right) \ , 
\label{2-16}\\
& &{\wtilde R}^{1,-1}(m;r,k)=
{\wtilde S}_{2m-2k+2}^{2m-2r+1} \ . 
\quad
\left(
\begin{array}{l}
r=1,2,\cdots, m-1, \\
k=r+1, r+2,\cdots, m
\end{array}
\right)
\label{2-17}
\eeq
The forms (\ref{2-13})$\sim$ (\ref{2-17}) are also for ${\wtilde S}^p$ and ${\wtilde S}_q^p$\ ($p>q)$. 
The expressions ($\beta$) and ($\epsilon$) give us scalars in the form 
\beq\label{2-18}
{\wtilde \Sigma}^{0,0}(m;r,k)=\sum_{\nu=\pm 1,0}(-)^\nu {\wtilde R}^{1,\nu}(m;r,k){\wtilde R}^{1,-\nu}(m;r,k)\ . 
\eeq
Hermitian conjugate of ${\wtilde R}^{\mu,\nu}(m;r,k)$ is denoted as 
\beq\label{2-19}
\left({\wtilde R}^{\mu,\nu}(m;r,k)\right)^*={\wtilde R}_{\mu,\nu}(m;r,k)\ . 
\eeq
For the present, the type ${\wtilde S}_p^p$ is re-formed as follows: 
\beq
& &{\rm (i)}\ \ \ n=2m ; \nonumber\\
& &\qquad
{\wtilde P}_0(m;r)=
\left\{
\begin{array}{ll}
{\displaystyle \frac{1}{2}\left({\wtilde S}_{2m-2r+1}^{2m-2r+1}+{\wtilde S}_{2m-2r}^{2m-2r}\right)} \ , & (r=1,\ 2, \cdots ,\ m-1) \\
{\displaystyle \frac{1}{2}{\wtilde S}_{1}^{1}} \ , & (r=m) 
\end{array}\right.
\label{2-20}\\
& &{\rm (ii)}\ \ n=2m+1 ; \nonumber\\
& &\qquad
{\wtilde P}_0(m;r)=
\frac{1}{2}\left({\wtilde S}_{2m-2r+2}^{2m-2r+2}+{\wtilde S}_{2m-2r+1}^{2m-2r+1}\right) \ . \quad (r=1,\ 2, \cdots ,\ m-1,\ m) 
\label{2-21}
\eeq
The operators ${\wtilde P}_0(m;r)$ should be compared with ${\wtilde S}_0(m;r)$. 
It is noted that we have the relation ${\wtilde P}_0(m;m)={\wtilde S}_0(m;m)$ in the case (i) and 
${\wtilde P}_0(m;r)\ (r=1,\ 2, \cdots ,\ m-1,\ m)$ are not scalars, but in the case (ii), 
${\wtilde P}_0(m;r)\ (r=1,\ 2, \cdots ,\ m-1,\ m)$ 
are scalars. 
Later, we will contact with this point.

Total number of ${\wtilde S}^p(n)$ $(p=1,\ 2,\cdots, \ n-1)$ and ${\wtilde S}_q^p(n)$ $(p>q=1,\ 2, \cdots ,\ n-2)$ is 
equal to 
\beq\label{2-22}
(n-1)+\frac{(n^2-1)-(n-1)}{2}=\frac{1}{2}n(n-1)\ . 
\eeq
On the other hand, total number of ${\wtilde S}_+(m;r)$, ${\wtilde R}^{0,0}(m;r,k)$ and 
${\wtilde R}^{1,\nu}(m;r,k)$ $(\nu=\pm 1, 0)$ in the case with $n=2m$ is equal to 
\beq\label{2-23}
m+\sum_{r=1}^{m-1}(m-r)+3\sum_{r=1}^{m-1}(m-r)=m(2m-1)=\frac{1}{2}n(n-1)\ . 
\eeq
In the case with $n=2m+1$, total number of ${\wtilde S}_+(m;r)$, ${\wtilde R}^{0,0}(m;r,k)$, ${\wtilde R}^{\frac{1}{2},\nu}(m;r)$ $(\nu=\pm 1/2)$ 
and ${\wtilde R}^{1,\nu}(m;r,k)$ $(\nu=\pm 1, 0)$ is equal to the relation (\ref{2-22}):
\beq\label{2-24}
m+\sum_{r=1}^{m-1}(m-r)+2m+3\sum_{r=1}^{m-1}(m-r)
=m(2m+1)=\frac{1}{2}n(n-1)\ .
\eeq
Total number of ${\wtilde S}_p^p(n)$ ($p=1,\ 2, \cdots ,\ n-1)$ is equal to $(n-1)$. 
In the case with $n=2m$, total number of ${\wtilde S}_0(m;r)$ and ${\wtilde P}_0(m;r)$ is equal to 
\beq\label{2-25}
m+m=2m=n\ .
\eeq
We have the difference $n-(n-1)=1$. 
The reason why we have this difference comes from the double counting for $(1/2){\wtilde S}_1^1$. 
Later, we will show that this double counting does not give us any trouble. 
In the case with $n=2m+1$, total number of ${\wtilde S}_0(m;r)$ and ${\wtilde P}_0(m;r)$ is equal to 
\beq\label{2-26}
m+m=2m=n-1\ .
\eeq
Thus, we re-formed the $su(n)$-generators in the Lipkin model in the form 
of the spherical tensor representation for the $su(2)$-algebra $({\wtilde S}_{\pm,0}(m))$.

Finally, we give a comment. 
We presented a possible re-formation of the original representation of the $su(n)$-Lipkin model. 
Concerning the $su(2)$-subalgebra, it starts in the expressions (\ref{2-3}) and (\ref{2-4}). 
However, it must be noted that these forms are not unique. 
If the indices $i=1,2,\cdots ,n-1$ are changed from the forms (\ref{2-3}) and (\ref{2-4}). we obtain 
various forms. 
For example, in the case with $n=2m=4$, 
${\wtilde S}_{\pm,0}(m;r)$ in the relation (\ref{2-3}) can be expressed as 
\bsub\label{2-27}
\beq
& &{\wtilde S}_+(2;1)={\wtilde S}_2^3\ , \quad
{\wtilde S}_-(2;1)={\wtilde S}_3^2\ , \quad
{\wtilde S}_0(2;1)=\frac{1}{2}\left({\wtilde S}_3^3-{\wtilde S}_2^2\right)\ , 
\label{2-27a}\\
& &{\wtilde S}_+(2;2)={\wtilde S}^1\ , \quad
{\wtilde S}_-(2;2)={\wtilde S}_1\ , \quad
{\wtilde S}_0(2;2)=\frac{1}{2}{\wtilde S}_1^1\ , 
\label{2-27b}
\eeq
\esub
For the form (\ref{2-27}), we exchange the indices $i=1$ and 2 and, then, they becomes
\bsub\label{2-28}
\beq
& &{\wtilde S}_+(2;1)={\wtilde S}_1^3\ , \quad
{\wtilde S}_-(2;1)={\wtilde S}_3^1\ , \quad
{\wtilde S}_0(2;1)=\frac{1}{2}\left({\wtilde S}_3^3-{\wtilde S}_1^1\right)\ , 
\label{2-28a}\\
& &{\wtilde S}_+(2;2)={\wtilde S}^2\ , \quad
{\wtilde S}_-(2;2)={\wtilde S}_2\ , \quad
{\wtilde S}_0(2;2)=\frac{1}{2}{\wtilde S}_2^2\ , 
\label{2-28b}
\eeq
\esub
The form (\ref{2-28}) also obeys the $su(2)$-algebra.

\setcounter{equation}{0}
\section{The Casimir operator in the present representation and its diagonalization}

In this section, mainly we will discuss the Casimir operator, ${\wtilde \Gamma}_{su(n)}$, 
shown in the relation (I.2.4):
\beq\label{3-1}
& &{\wtilde \Gamma}_{su(n)}=
\frac{1}{2}\left[
\sum_{p=1}^{n-1}\left[\ {\wtilde S}^p\ , \ {\wtilde S}_p\ \right]_+
+\sum_{p=2}^{n-1}\sum_{q=1}^{p-1}\left[\ {\wtilde S}_q^p\ , \ {\wtilde S}_p^q\ \right]_+
+\sum_{p=1}^{n-1}\left({\wtilde S}_p^p\right)^2
-\frac{1}{n}\left(\sum_{p=1}^{n-1}{\wtilde S}_p^p\right)^2\right]\ . \nonumber\\
& &
\eeq
Here, $[\ {\wtilde A}\ , \ {\wtilde B}\ ]_+={\wtilde A}{\wtilde B}+{\wtilde B}{\wtilde A}$. 
We can see in the form (\ref{3-1}) that ${\wtilde \Gamma}_{su(n)}$, 
which is of the quadratic form with respect to the $su(n)$-generators, 
is not diagonal. 
Then, we will consider the diagonalization. 
%as a quadratic form 
%with respect to the $su(n)$-generators in the original representation is not diagonalized. 
The operator ${\wtilde \Gamma}_{su(n)}$ can be rewritten as 
\beq
& &{\wtilde \Gamma}_{su(n)}=\sum_{r=1}^{m}{\wtilde {\mib S}}(m;r)^2+
{\wtilde {\mib \Gamma}}_m+{\wtilde {\mib \Gamma}}_m^{(2)}\ , 
\label{3-2}\\
& &{\wtilde {\mib S}}(m;r)^2=
\frac{1}{2}\left[\ {\wtilde S}_+(m;r)\ , \ {\wtilde S}_-(m;r)\ \right]_+
+{\wtilde S}_0(m;r)^2\ . 
\label{3-3}
\eeq
The part ${\wtilde {\mib S}}(m;r)^2$ denotes the Casimir operator of the $su(2)$-subalgebra 
$({\wtilde S}_{\pm,0}(m;r))$ and ${\wtilde {\mib \Gamma}}_m$ is given separately in the cases with 
$n=2m$ and $2m+1$:
\bsub\label{3-4}
\beq
& &{\rm (i)}\ \ \ n=2m;
\nonumber\\
& &\quad
{\wtilde {\mib \Gamma}}_m=\frac{1}{2}\sum_{r=1}^{m-1}\sum_{k=r+1}^m
\Biggl(\left[\ {\wtilde R}^{0,0}(m;r,k)\ , \ {\wtilde R}_{0,0}(m;r,k)\ \right]_+
\nonumber\\
& &\qquad\qquad\qquad\qquad\ \ 
+\sum_{\nu=\pm 1,0}
\left[\ {\wtilde R}^{1,\nu}(m;r,k)\ , \ {\wtilde R}_{1,\nu}(m;r,k)\ \right]_+\Biggl)\ , 
\label{3-4a}\\
& &{\rm (ii)}\ \ n=2m+1;
\nonumber\\
& &\quad
{\wtilde {\mib \Gamma}}_m=\frac{1}{2}\sum_{r=1}^{m-1}\sum_{k=r+1}^m
\Biggl(\left[\ {\wtilde R}^{0,0}(m;r,k)\ , \ {\wtilde R}_{0,0}(m;r,k)\ \right]_+
\nonumber\\
& &\qquad\qquad\qquad\qquad\ \ 
+\sum_{\nu=\pm 1,0}
\left[\ {\wtilde R}^{1,\nu}(m;r,k)\ , \ {\wtilde R}_{1,\nu}(m;r,k)\ \right]_+\Biggl)
\nonumber\\
& &\qquad\ \ \ 
+\frac{1}{2}\sum_{r=1}^m\sum_{\nu=\pm 1/2}
\left[\ {\wtilde R}^{\frac{1}{2},\nu}(m;r)\ , \ {\wtilde R}_{\frac{1}{2},\nu}(m;r)\ \right]_+\ .
\label{3-4b}
\eeq
\esub
The part ${\wtilde {\mib \Gamma}}_m^{(2)}$ is expressed as 
\beq\label{3-5}
{\wtilde {\mib \Gamma}}_m^{(2)}
=\sum_{r=1}^{m}\left({\wtilde P}_0(m;r)\right)^2
-\frac{2}{n}\left(\sum_{r=1}^{m}{\wtilde P}_0(m;r)\right)^2\ . 
\eeq
The operators ${\wtilde {\mib \Gamma}}_m$ and ${\wtilde {\mib \Gamma}}_m^{(2)}$ are 
of the quadratic. 
However, ${\wtilde {\mib \Gamma}}_m$ is diagonalized, 
but ${\wtilde {\mib \Gamma}}_m^{(2)}$ is not diagonalized. 
The above is our re-formation of ${\wtilde \Gamma}_{su(n)}$.

The operator ${\wtilde {\mib \Gamma}}_m^{(2)}$ can be diagonalized in the following: 
\beq\label{3-6}
{\wtilde {\mib \Gamma}}_m^{(2)}
=\sum_{r=1}^{m-1}\left({\wtilde Q}_0(m;r)\right)^2
+\left(\frac{n-2m}{n}\right)\left({\wtilde Q}_0(m;m)\right)^2\ , 
\eeq
i.e., 
\bsub\label{3-7}
\beq
& &{\rm (i)}\ \ \ n=2m;\quad\qquad
{\wtilde {\mib \Gamma}}_{m=1}^{(2)}=0\ , \qquad
{\wtilde {\mib \Gamma}}_{m\geq 2}^{(2)}=\sum_{r=1}^{m-1}\left({\wtilde Q}_0(m;r)\right)^2\ , 
\label{3-7a}\\
& &{\rm (ii)}\ \ n=2m+1;\quad
{\wtilde {\mib \Gamma}}_{m=1}^{(2)}=\frac{1}{3}\left({\wtilde Q}_0(1;1)\right)^2, \nonumber\\
& &\qquad\qquad\qquad\qquad\ \ 
{\wtilde {\mib \Gamma}}_{m\geq 2}^{(2)}=\sum_{r=1}^{m-1}
\left({\wtilde Q}_0(m;r)\right)^2+\frac{1}{2m+1}\left({\wtilde Q}_0(m;m)\right)^2. \qquad\ \ 
\label{3-7b}
\eeq
\esub
It should be noted that the second term on the right-hand side of the 
relation (\ref{3-6}) disappears in the case with $n=2m$. 
The operator ${\wtilde Q}_0(m;r)$ is obtained 
from ${\wtilde P}_0(m;r)$ by the orthogonal transformation 
\bsub\label{3-8}
\beq
{\wtilde Q}_0(m;r)=\sum_{r'=1}^m C_{r,r'}(m){\wtilde P}_0(m;r')\ , 
\label{3-8a}
\eeq
conversely,
\beq
{\wtilde P}_0(m;r)=\sum_{r'=1}^m C_{r',r}(m){\wtilde Q}_0(m;r')\ . 
\label{3-8b}
\eeq
\esub
Here, $(C_{r',r}(m))$ denotes an orthogonal matrix:
\beq\label{3-9}
C_{r'=m,r}(m)=
\frac{1}{\sqrt{m}}\ , \quad
\sum_{r''=1}^m C_{r,r''}(m)C_{r',r''}(m)
=\sum_{r''=1}^m C_{r'',r}(m)C_{r'',r'}(m)=\delta_{r,r'}.\qquad
\eeq
In this paper, we will adopt the following form for $C_{r',r}(m)$:
\beq\label{3-10}
C_{r',r}(m)
=\biggl\langle\frac{m-1}{2},r-\frac{m+1}{2},\frac{m-1}{2},-\left(r-\frac{m+1}{2}\right)
\biggl| m-r',0\biggl\rangle (-)^{r-r'}\ . \qquad
\eeq
Here, $C_{r',r}(m)$ except for the phase factor $(-)^{r-r'}$ denotes the 
Clebsch-Gordan coefficient of the type 
$\bra{j_1,m_1,j_1,-m}j_2,0\rangle$ and it is given as 
\setcounter{equation}{9}
\bsub
\beq
& &C_{r',r}(m)=(-)^{m-r'}
\left(\sqrt{D_{m,r'}}\right)^{-1}E_{r',r}(m)\ , 
\label{3-10a}\\
& &D_{m,r'}=\frac{\prod_{\mu=0}^{m-r'}(m+\mu)(m-\mu)}{m(2(m-r')+1)}\ , 
\label{3-10b}\\
& &E_{r',r}(m)
=\sum_s(-)^s\frac{(r-1+s)!(2m-r'-r-s)!}{s!(m-r-s)!(m-r'-s)!(r'+r-m-1+s)!}\ , 
\label{3-10c}\\
& &F_{r',r}(m)=(-)^{m-r'}\left(D_{m,r'}\right)^{-1}E_{r',r}(m)\ . 
\label{3-10d}
\eeq
\esub
Later, $F_{r',r}(m)$ will be used. 
For example, we have 
\bsub\label{3-11}
\beq
& &
C_{r'=m,r}(m)=\frac{1}{\sqrt{m}}\ , \qquad
D_{m,r'=m}=m\ , \qquad
E_{r'=m,r}(m)=1\ , 
\label{3-11a}\\
& &
C_{r'=m-1,r}(m)=-\sqrt{\frac{3}{m(m+1)(m-1)}}(2r-(m+1))\ , 
\nonumber\\
& &D_{m,r'=m-1}=\frac{m(m+1)(m-1)}{3}\ , \qquad
E_{r'=m-1,r}(m)=2r-(m+1)\ . 
\label{3-11b}
\eeq
\esub
The relations (\ref{3-8}) and (\ref{3-9}) lead us to the expression (\ref{3-6}). 
We note that ${\wtilde Q}_0(m;r)$ for $r=1,\ 2,\cdots ,\ m-1$ are scalar in the case with $n=2m$. 
Later, it will be shown. 
In the case with $n=2m$, ${\wtilde Q}_0(m;m)$ disappears in ${\wtilde {\mib \Gamma}}_m^{(2)}$, 
because of the factor $(n-2m)$ in the relation (\ref{3-6}). 
In this case, in spite of the relation 
${\wtilde P}_0(m;m)={\wtilde S}_0(m;m)$, we have treated these two as if 
they are independent of each other. 
As was already mentioned in the relation (\ref{2-25}), total number of 
${\wtilde P}_0(m;r)$, i.e., ${\wtilde Q}_0(m;r)$ becomes equal to that of 
${\wtilde S}_0(m;r)$.
Originally, total number of ${\wtilde S}_p^p(n)$ is equal to $n-1$, i.e., 
$2m-1$ and in this number, the integer $m$ is fixed as total number of ${\wtilde S}_0(m;r)$. 
Therefore, total number of ${\wtilde Q}_0(m;r)$ should be reduced to $m-1$. 
We can see in the relation (\ref{3-7a}) that ${\wtilde Q}_0(m;m)$ disappears, in other word, only 
${\wtilde Q}_0(m;r)$ $(r=1,\ 2,\cdots ,\ m-1)$ appear. 
The case with $n=2m+1$ does not contain such a problem, because of $n=2m+1$, 
i.e., $n-1=2m$.

Now, let us derive the formulae for obtaining the eigenvalues of ${\wtilde \Gamma}_{su(n)}$. 
For this aim, we must rewrite $(1/2)\cdot
[\ {\wtilde S}_+(m;r)\ , \ {\wtilde S}_-(m;r)\ ]_+$ and 
other terms of the anti-commutators to the forms 
${\wtilde S}_+(m;r){\wtilde S}_-(m;r)$ and others. 
As the result, the linear terms for the generators appear. 
The part $\sum_{r=1}^m{\wtilde {\mib S}}(m;r)^2$ shown in the relation 
(\ref{3-3}) is simple: 
\beq\label{3-12}
{\rm the\ linear\ term}=-\sum_{r=1}^{m}{\wtilde S}_0(m;r)\ . 
\eeq
The linear terms coming from the relation (\ref{3-4}), which we denote as 
${\wtilde {\mib \Gamma}}_m^{(1)}$ can be expressed in the form 
\beq\label{3-13}
{\wtilde {\mib \Gamma}}_m^{(1)}
&=&2\sum_{r=1}^m\left(2r-\left(\frac{n}{2}+1\right)\right){\wtilde P}_0(m;r)
\nonumber\\
&=&2\sum_{r=1}^m\left(2r-(m+1)\right){\wtilde P}_0(m;r)
-(n-2m)\sum_{r=1}^m{\wtilde P}_0(m;r)\ . 
\eeq
With the use of the relations (\ref{3-8}) and (\ref{3-11}), 
${\wtilde {\mib \Gamma}}_m^{(1)}$ can be expressed as 
\beq\label{3-14}
{\wtilde {\mib \Gamma}}_m^{(1)}
=-2\sqrt{D_{m,m-1}}{\wtilde Q}_0(m,m-1)-(n-2m)\sqrt{m}{\wtilde Q}_0(m;m)\ . 
\eeq
We can see that in the case with $n=2m$, ${\wtilde Q}_0(m;m)$ disappears also in 
${\wtilde {\mib \Gamma}}_m^{(1)}$.

The sum ${\wtilde {\mib \Gamma}}_m^{(2)}+{\wtilde {\mib \Gamma}}_m^{(1)}$ 
$(={\wtilde {\mib \Gamma}}_m^{(0)})$ can be expressed in the following form: 
\beq\label{3-15}
{\wtilde {\mib \Gamma}}_m^{(0)}
&=&\sum_{r=1}^{m-1}{\wtilde Q}_0(m;r)
\left({\wtilde Q}_0(m;r)-2\sqrt{D_{m,r}}\delta_{r,m-1}\right)
\nonumber\\
& &+\left(\frac{n-2m}{n}\right){\wtilde Q}_0(m;m)
\left({\wtilde Q}_0(m;m)-n\sqrt{m}\right)\ .
\eeq
By changing the scale of ${\wtilde Q}_0(m;r)$, we can eliminate the irrational 
numbers contained in ${\wtilde Q}_0(m;r)$:
\beq\label{3-16}
{\wtilde {\mib \Gamma}}_m^{(0)}
&=&
\frac{1}{4}\sum_{r=1}^{m-1}D_{m,r}{\wtilde R}_0(m;r)\left(
{\wtilde R}_0(m;r)-4\delta_{r,m-1}\right)
\nonumber\\
& &+(n-2m)\cdot\frac{1}{nm}{\wtilde R}_0(m;m)\left({\wtilde R}_0(m;m)-nm\right)\ , 
\qquad\qquad\qquad\qquad\qquad\qquad
\eeq
\bsub\label{3-17}
\beq
{\wtilde R}_0(m;r)
&=&2\left(\sqrt{D_{m,r}}\right)^{-1}{\wtilde Q}_0(m;r)
\nonumber\\
&=&2\sum_{r'=1}^m
F_{r,r'}(m){\wtilde P}_0(m;r')\ , \quad (r=1,\ 2,\cdots ,\ m-1)
\qquad\quad
\label{3-17a}\\
{\wtilde R}_0(m;m)&=&
\sqrt{m}{\wtilde Q}_0(m;m)=\sum_{r'=1}^m{\wtilde P}_0(m;r')\ . 
\label{3-17b}
\eeq
\esub
The linear terms indicate the quantum fluctuations. 
We see that the fluctuations appear only at the points $r=m$ and $m-1$. 
If the orthogonal matrix (\ref{3-10}) is not adopted, they may spread 
over various points of $r$. 
In the case with $n=2m$, ${\wtilde P}_0(m;r)$ $(r=1,\ 2,\cdots ,\ m-1,\ m)$ 
is not scalar. 
But, it can be easily proved that ${\wtilde Q}_0(m;r)$, i.e., 
${\wtilde R}_0(m;r)$ $(r=1,\ 2,\cdots ,\ m-1)$ is scalar and we see that, 
in ${\wtilde {\mib \Gamma}}_m^{(0)}$, ${\wtilde R}_0(m;m)$ in the case with $n=2m$ disappears. 
With the use of the results in (I), we can calculate the eigenvalues of 
${\wtilde R}_0(m;r)$ for a given minimum weight state.

Let us show that ${\wtilde Q}_0(m;r)$ for $r=1,\ 2,\cdots,\ m-1$ in the case with 
$n=2m$ is scalar. 
For example, we have 
\beq\label{3-18}
[\ {\wtilde S}_+(m)\ , \ {\wtilde P}_0(m;r)\ ]=-{\wtilde S}^1\ . \qquad (r=1,\ 2,\cdots ,\ m-1,\ m)
\eeq
Then, the relation (\ref{3-18}) with (\ref{3-9}) and (\ref{3-11a}) leads us to 
\beq\label{3-19}
[\ {\wtilde S}_+(m)\ , \ {\wtilde Q}_0(m;r)\ ]
&=&
\sum_{r'=1}^m C_{r,r'}(m)
[\ {\wtilde S}_+(m)\ , \ {\wtilde P}_0(m)\ ]
\nonumber\\
&=&
-\left(\sum_{r'=1}^m C_{r,r'}(m)\right){\wtilde S}^1
=-\delta_{r,m}\sqrt{m}\ {\wtilde S}^1\ . 
\eeq
In the case with $n=2m+1$ for $r=1,\ 2,\cdots ,\ m-1,\ m$, we have 
\beq\label{3-20}
[\ {\wtilde S}_+(m)\ , \ {\wtilde P}_0(m;r)\ ]=0\ , \quad
{\rm i.e.,}\quad
[\ {\wtilde S}_+(m)\ , \ {\wtilde Q}_0(m;r)\ ]=0\ .
\eeq
Other cases with ${\wtilde S}_-(m)$ and ${\wtilde S}_0(m)$ are also 
in the same situation as the above.

Further, for $r=1,\ 2,\cdots ,\ m-1$ in the case with $n=2m$, the following relation is derived: 
\bsub\label{3-21}
\beq
{\wtilde Q}_0(m;r)=\sum_{r'=1}^{m-1}C_{r,r'}(m){\wtilde {\cal P}}_0(m;r')\ , 
\label{3-21a}
\eeq
conversely,
\beq
{\wtilde {\cal P}}_0(m;r)-\frac{1}{m}\sum_{r'=1}^{m-1}{\wtilde {\cal P}}_0(m;r')
=\sum_{r'=1}^{m-1}C_{r',r}(m){\wtilde Q}_0(m;r')\ . 
\label{3-21b}
\eeq
\esub
Here, ${\wtilde {\cal P}}_0(m;r)$ is defined as 
\beq\label{3-22}
{\wtilde {\cal P}}_0(m;r)
={\wtilde P}_0(m;r)-{\wtilde P}_0(m;m)
=\frac{1}{2}\left({\wtilde S}_{2m-2r+1}^{2m-2r+1}+{\wtilde S}_{2m-2r}^{2m-2r}
-{\wtilde S}_1^1\right)\ .
\eeq
Therefore, in place of using ${\wtilde Q}_0(m;r)$, it may be 
permitted also to use ${\wtilde {\cal P}}_0(m;r)$. 
In parallel to the case with $n=2m$, we also use ${\wtilde {\cal P}}_0(m;r)$ 
in the case with $n=2m+1$: 
\beq\label{3-23}
& &{\wtilde {\cal P}}_0(m;r)
={\wtilde P}_0(m;r)
=\frac{1}{2}\left({\wtilde S}_{2m-2r+2}^{2m-2r+2}+{\wtilde S}_{2m-2r+1}^{2m-2r+1}
\right)\ . \quad (r=1,\ 2,\cdots ,\ m-1,\ m)\nonumber\\
& &
\eeq
However, the relation (\ref{3-21}) does not hold in the case with $n=2m+1$.

In (I), we showed an example of the Hamiltonian: 
\beq\label{3-24}
{\wtilde H}(n)={\wtilde H}_0(n)+{\wtilde H}_1(n)\ . 
\eeq
The term ${\wtilde H}_0(n)$ is the Hamiltonian of individual levels with energies $\epsilon_p$: 
\bsub\label{3-25}
\beq
& &{\wtilde H}_0(n)=\sum_{p=0}^{n-1}
\epsilon_p{\wtilde N}_p(n)=\sum_{p=1}^{n-1}\epsilon_p{\wtilde S}_p^p(n)\ , 
\label{3-25a}\\
& &\sum_{p=0}^{n-1}\epsilon_p=0\ , \quad \epsilon_0 \leq \epsilon_1\leq \cdots \leq \epsilon_{n-1}\ .
\label{3-25b}
\eeq
\esub
The part ${\wtilde H}_1(n)$ is the interaction term: 
\beq\label{3-26}
{\wtilde H}_1(n)=
-G\sum_{p=1}^{n-1}\left[
\left({\wtilde S}^p(n)\right)^2+\left({\wtilde S}_p(n)\right)^2\right]\ . 
\quad
(G:{\rm coupling\ constant})
\eeq
We can rewrite ${\wtilde H}(n)$ in our new representation. 
The part ${\wtilde H}_0(n)$ can be rewritten in the form 
given below. 
In the case with $n=2m$, the relations (\ref{2-3}) and (\ref{2-26}) give us 
\bsub\label{3-27}
\beq
& &{\wtilde S}_{2m-2r+1}^{2m-2r+1}
={\wtilde P}_0(m;r)+{\wtilde S}_0(m;r)\ , \quad
{\wtilde S}_{2m-2r}^{2m-2r}={\wtilde P}_0(m;r)-{\wtilde S}_0(m;r)\ , 
\nonumber\\
& &\qquad\qquad\qquad\qquad\qquad\qquad\qquad\qquad\qquad\qquad
(r=1,2,\cdots ,m-1)
\label{3-27a}\\
& &{\wtilde S}_1^1={\wtilde P}_0(m;m)+{\wtilde S}_0(m;m)\ . \quad
\left(r=m,\ {\wtilde P}_0(m;m)={\wtilde S}_0(m;m)\right)
\label{3-27b}
\eeq
\esub
Then, ${\wtilde H}_0(n)$ can be expressed in the form 
\beq\label{3-28}
{\wtilde H}_0(n)&=&
\sum_{r=1}^m (\epsilon_{2m-2r+1}+\epsilon_{2m-2r})
{\wtilde P}_0(m;r)
+\sum_{r=1}^m (\epsilon_{2m-2r+1}-\epsilon_{2m-2r}){\wtilde S}_0(m;r)
\nonumber\\
&=&
\sum_{r=1}^{m-1}(\epsilon_{2m-2r+1}+\epsilon_{2m-2r}){\wtilde {\cal P}}_0(m;r)
+\sum_{r=1}^m (\epsilon_{2m-2r+1}-\epsilon_{2m-2r}){\wtilde S}_0(m;r)\ . \ \ \qquad 
\eeq
For the above rewriting, the relation (\ref{3-25b}) is used. 
In the case with $n=2m+1$, we have
\beq\label{3-29}
& &{\wtilde S}_{2m-2r+2}^{2m-2r+2}={\wtilde P}_0(m;r)+{\wtilde S}_0(m;r)\ , \qquad
{\wtilde S}_{2m-2r+1}^{2m-2r+1}={\wtilde P}_0(m;r)-{\wtilde S}_0(m;r)\ . 
\nonumber\\
& &\qquad\qquad\qquad\qquad\qquad\qquad\qquad\qquad\qquad
(r=1,\ 2,\cdots ,\ m-1,\ m)
\eeq
Then, we obtain 
\beq\label{3-30}
& &{\wtilde H}_0(n)=
\sum_{r=1}^m (\epsilon_{2m-2r+2}+\epsilon_{2m-2r+1}){\wtilde {\cal P}}_0(m;r)
+\sum_{r=1}^m (\epsilon_{2m-2r+2}-\epsilon_{2m-2r+1}){\wtilde S}_0(m;r)\ . \nonumber\\
& &
\eeq

The term ${\wtilde H}_1(n)$ is rather simple. 
In the case with $n=2m$, we have 
\beq\label{3-31}
& &{\wtilde S}^{2m-2r+1}=-{\wtilde R}^{1,1}(m;r,m)\ , 
\nonumber\\
& &{\wtilde S}^{2m-2r}=\frac{1}{\sqrt{2}}\left(
{\wtilde R}^{0,0}(m;r,m)-{\wtilde R}^{1,0}(m;r,m)\right)\ , \quad
(r=1,\ 2,\cdots ,\ m-1)
\nonumber\\
& &{\wtilde S}_+(m;m)={\wtilde S}^1\ .
\eeq
In the case with $n=2m+1$, we obtain 
\beq\label{3-32}
& &{\wtilde S}^{2m-2r+2}={\wtilde R}^{\frac{1}{2},\frac{1}{2}}(m;r)\ , 
\nonumber\\
& &{\wtilde S}^{2m-2r+1}={\wtilde R}^{\frac{1}{2},-\frac{1}{2}}(m;r)\ . 
\quad
(r=1,\ 2,\cdots ,\ m-1,\ m)
\eeq
Then, we have 
\beq
& &\sum_{p=1}^{n-1}\left({\wtilde S}^p\right)^2
=\sum_{r=1}^{m-1}\left[
\left({\wtilde R}^{1,1}(m;r,m)\right)^2
+\frac{1}{2}\left({\wtilde R}^{0,0}(m;r,m)-{\wtilde R}^{1,0}(m;r,m)\right)^2\right]
\nonumber\\
& &\qquad\qquad\qquad
+\left({\wtilde S}_+(m;m)\right)^2\ , 
%\nonumber\\
%& &\qquad\qquad\qquad\qquad\qquad\qquad\qquad\qquad\qquad
\qquad\qquad\qquad
({\rm the\ case\ with}\ n=2m)
\label{3-33}\\
& &\sum_{p=1}^{n-1}\left({\wtilde S}^p\right)^2
=\sum_{r=1}^{m}\left[
\left({\wtilde R}^{\frac{1}{2},\frac{1}{2}}(m;r)\right)^2
+\left({\wtilde R}^{\frac{1}{2},-\frac{1}{2}}(m;r)\right)^2\right]\ .
\nonumber\\
& &\qquad\qquad\qquad\qquad\qquad\qquad\qquad\qquad\qquad
\quad ({\rm the\ case\ with}\ n=2m+1)
\label{3-34}
\eeq
The part ${\wtilde H}_1(n)$ can be given by substituting the relations (\ref{3-33}) and 
(\ref{3-34}) and their hermitian conjugate into the relation (\ref{3-26}).

\setcounter{equation}{0}
\section{A possible scheme for constructing the linearly independent basis}

In (I), we investigated the structure of the minimum weight states of 
the $su(n)$-Lipkin model. 
In {\bf 6} of (I), we sketched the minimum weight states in the representation 
developed in the present paper (II) for the cases with $n=2,\ 3,\ 4$ and 5. 
Although our final aim of this section is to present our idea for constructing the linearly independent basis, the first task is to generalize the above four cases to the case with 
arbitrary $n$.

Let the eigenvalues of ${\wtilde S}_0(m;r)$ and ${\wtilde R}_0(m;r)$ for 
the minimum weight state be denoted as $-\sigma^r$ and $-\rho^r$, respectively. 
Here, ${\wtilde S}_0(m;r)$ is given in the relation (\ref{2-3}) for the case with $n=2m$ and 
(\ref{2-4}) for the case with $n=2m+1$: 
\bsub\label{4-1}
\beq
& &{\rm (i)}\ \ \ n=2m,\nonumber\\
& &\qquad
{\wtilde S}_0(m;r)=
\left\{
\begin{array}{ll}
{\displaystyle \frac{1}{2}\left({\wtilde S}_{2m-2r+1}^{2m-2r+1}-{\wtilde S}_{2m-2r}^{2m-2r}
\right)}\ , & (r=1,\ 2, \cdots,\ m-1) \\
 & \\
{\displaystyle \frac{1}{2}{\wtilde S}_1^1}\ , & 
(r=m) 
\end{array}\right.
\label{4-1a}\\
& &{\rm (ii)}\ \ n=2m+1,\nonumber\\
& &\qquad
{\wtilde S}_0(m;r)=\frac{1}{2}\left({\wtilde S}_{2m-2r+2}^{2m-2r+2}-{\wtilde S}_{2m-2r+1}^{2m-2r+1}
\right)\ . \quad 
(r=1,\ 2, \cdots ,\ m-1,\ m)\quad
\label{4-1b}
\eeq
\esub
On the other hand, the relations (\ref{3-17}) and (\ref{3-21}) with (\ref{3-10}) gives us the operators ${\wtilde R}_0(m;r)$: 
\bsub\label{4-2}
\beq
& &{\rm (i)}\ \ \ n=2m,\nonumber\\
& &\qquad
{\wtilde R}_0(m;r)=2\sum_{r'=1}^{m-1}F_{r,r'}(m)\cdot{\wtilde {\cal P}}_0(m;r') 
%\frac{1}{2}
%\left({\wtilde S}_{2m-2r'+1}^{2m-2r'+1}+{\wtilde S}_{2m-2r'}^{2m-2r'}-{\wtilde S}_1^1\right)
\ , 
%\nonumber\\
%& &\qquad\qquad\qquad\qquad\qquad\qquad\qquad\qquad\qquad\qquad\qquad\qquad
\qquad
(r=1,\ 2, \cdots ,\ m-1)
\label{4-2a}\\
& &{\rm (ii)}\ \ n=2m+1,\nonumber\\
& &\qquad
{\wtilde R}_0(m;r)=
\left\{
\begin{array}{l}
{\displaystyle 2\sum_{r'=1}^{m}F_{r,r'}(m)
{\wtilde {\cal P}}_0(m;r')}
%\cdot\frac{1}{2}\left({\wtilde S}_{2m-2r'+2}^{2m-2r'+2}+{\wtilde S}_{2m-2r'+1}^{2m-2r'+1}}
%\right)}
\ , 
%\\
%\qquad\qquad\qquad\qquad\qquad\qquad\qquad
\qquad
(r=1,\ 2, \cdots,\ m-1) \\
{\displaystyle \sum_{r'=1}^m{\wtilde {\cal P}}_0(m;r')}
%\frac{1}{2}\left({\wtilde S}_{2m-2r'+2}^{2m-2r'+2}
%+{\wtilde S}_{2m-2r'+1}^{2m-2r'+1}\right)}
\ . 
\qquad\qquad\qquad
(r=m) 
\end{array}\right.
\label{4-2b}
\eeq
\esub
Here, ${\wtilde {\cal P}}_0(m;r)$ is shown in the relations (\ref{3-22}) and (\ref{3-23}). 
We notice that there exists a difference between the case with $n=2m$ and with $n=2m+1$. 
For $\sigma^r$, the superscript $r$ is in the range $r=1,\ 2,\cdots,\ m-1,\ m$, 
for both cases. 
However, for $\rho^r$, the range of $r$ in the case with $n=2m$ is 
$r=1,\ 2,\cdots ,\ m-1$ and in the case with $n=2m+1$, $r=1,\ 2,\cdots ,\ m-1,\ m$. 
The reason has been already mentioned and also given in {\bf A}. 
Then, the minimum weight state $\ket{{\rm min}(n)}$ can be expressed formally in the form 
\beq\label{4-3}
\ket{{\rm min}(n)}
=\left\{
\begin{array}{ll}
\ket{N;{\vec{\rho}},\vec{\sigma}}\ , & (n=2m) \\
\ket{N;\vec{\rho},\rho_m,\vec{\sigma}}\ , & (n=2m+1)
\end{array}
\right.
\eeq
Here, we adopt the abbreviations 
\beq\label{4-4}
\vec{\rho}=\rho^1,\ \rho^2,\cdots ,\ \rho^{m-1}\ , \qquad
\vec{\sigma}=\sigma^1,\ \sigma^2,\cdots ,\ \sigma^{m-1},\ \sigma^m\ .
\eeq
As is clear from the relation (\ref{4-2}), it may be enough to obtain the eigenvalues of ${\wtilde {\cal P}}_0(m;r)$ $(r=1,2,\cdots ,m-1\ {\rm or}\ m)$. 
Then, we can calculate ${\vec \rho}$ and $\rho^m$.
We can treat ${\wtilde S}_+(m;r)$ in terms of operating ${\wtilde S}_+(m;\vec{\sigma}^0)$, 
which is defined in the following, on $\ket{{\rm min}(n)}$: 
\beq\label{4-5}
{\wtilde S}_+(m;\vec{\sigma}^0)=\prod_{r=1}^m
\left({\wtilde S}_+(m;r)\right)^{\sigma^0(r)}\ , 
\eeq
i.e., 
\bsub\label{4-6}
\beq
& &\ket{N;\vec{\rho},\vec{\sigma}^0}={\wtilde S}_+(m;\vec{\sigma}^0)\ket{N;\vec{\rho},\vec{\sigma}}
\ , \qquad\qquad\qquad
(n=2m)
\label{4-6a}\\
& &\ket{N;\vec{\rho},\rho_m,\vec{\sigma}^0}={\wtilde S}_+(m;\vec{\sigma}^0)
\ket{N;\vec{\rho},\rho^m,\vec{\sigma}}
\ . \qquad
(n=2m+1)
\label{4-6b}
\eeq
\esub
\beq\label{4-7}
\vec{\sigma}^0=\sigma^0(1),\ \sigma^0(2),\cdots ,\ \sigma^0(m)\ . 
\qquad\qquad\qquad\qquad\quad
\eeq
Since $[\ {\wtilde S}_+(m;r)\ , \ {\wtilde S}_+(m;r')\ ]=0$, the ordering 
of ${\wtilde S}_+(m;r)$ in ${\wtilde S}_+(m;{\vec\sigma}^0)$ is arbitrary. 
In the notation familiar to the $su(2)$-algebra, $\sigma^0(r)$ is expressed as 
\beq\label{4-8}
\sigma^0(r)=\sigma^r+\sigma_0^r\ .
\eeq
Of course, with the use of the angular momentum coupling rule, we are able to obtain 
the eigenstate of $({\wtilde S}_{\pm,0}(m))$. 
%The cases with $n=2,\ 3,\ 4$ and 5 have been shown in (I). 

Next, we consider some problems related to the quantum number $\rho^r$. 
We notice ${\wtilde R}^{0,0}(m;r,k)$ defined in the relations (\ref{2-9}) and 
(\ref{2-13}), their hermitian conjugates, ${\wtilde R}_{0,0}(m;r,k)$ and 
${\wtilde R}_0(m;r)$ introduced in the relation (\ref{3-17a}) form the $su(m)$-algebras 
for the cases with $n=2m$ and $n=2m+1$. 
The detail is given in {\bf A}. 
First, we show that the state (\ref{4-6}) is the minimum weight state of the $su(m)$-algebra. 
For this aim, the following relations for the cases with $n=2m$ and $2m+1$ are useful:
\bsub\label{4-9}
\beq
& &[\ {\wtilde R}_{0,0}(m;r,k)\ , \ {\wtilde S}_+(m;r')\ ]
=(\delta_{rr'}-\delta_{kr'}){\wtilde R}_{1,-1}(m;r,k)\ , 
\label{4-9a}\\
& &[\ {\wtilde R}_{1,-1}(m;r,k)\ , \ {\wtilde S}_+(m;r')\ ]=0\ , 
\label{4-9b}\\
& &{\wtilde R}_{1,-1}(m;r,k)\ket{{\rm min}(n)}=0\ , \qquad
{\wtilde R}_{0,0}(m;r,k)\ket{{\rm min}(n)}=0\ . 
\label{4-9c}
\eeq
\esub
The relation (\ref{4-9}) leads us to 
\bsub\label{4-10}
\beq
& &{\wtilde R}_{0,0}(m;r,k)\ket{N;\vec{\rho},\vec{\sigma}^0}=0\ , 
\label{4-10a}\\
& &{\wtilde R}_{0,0}(m;r,k)\ket{N;\vec{\rho},\rho^m,\vec{\sigma}^0}=0\ . 
\label{4-10b}
\eeq
\esub
Further, we have 
\beq\label{4-11}
[\ {\wtilde R}_{0}(m;r)\ , \ {\wtilde S}_+(m;r')\ ]=0\ ,\qquad
{\wtilde R}_0(m;r)\ket{{\rm min}(n)}=-\rho^r\ket{{\rm min}(n)}\ .  
\eeq
Then, we obtain 
\bsub\label{4-12}
\beq
& &{\wtilde R}_{0}(m;r)\ket{N;\vec{\rho},\vec{\sigma}^0}=-\rho^r\ket{N;\vec{\rho},\vec{\sigma}^0}
\ , 
\label{4-12a}\\
& &{\wtilde R}_{0}(m;r)\ket{N;\vec{\rho},\rho^m,\vec{\sigma}^0}=
-\rho^r\ket{N;\vec{\rho},\rho^m,\vec{\sigma}^0}\ . 
\label{4-12b}
\eeq
\esub
The relations (\ref{4-10}) and (\ref{4-12}) tell us that the state (\ref{4-6}) 
is the minimum weight state of the $su(m)$-algebra.

Since we get the minimum weight states of the present $su(m)$-algebra, 
we will consider ${\wtilde R}^{0,0}(m;r,k)$ operated on these states. 
First, we define the following raising operator: 
\beq
& &{\wtilde R}^{0,0}(m;{\vec \lambda}^0)=\prod_{r=1}^{m-1}
\left[\prod_{k=r+1}^{m}\left({\wtilde R}^{0,0}(m;r,k)\right)^{\lambda^0(r,k)}\right]\ , 
\label{4-13}\\
& &{\vec \lambda}^0=\lambda^0(1,2),\ \lambda^0(1,3),\cdots ,\ \lambda^0(1,m),\ \lambda^0(2,3),\ \lambda^0(2,4),
\cdots , \ \lambda^0(2,m),\cdots ,\ \lambda^0(m-1,m)\ . 
\nonumber\\
& &
\label{4-14}
\eeq
It should be noted that, since the operators ${\wtilde R}^{0,0}(m;r,k)$ are not mutually commuted, 
the ordering of their product is fixed beforehand. 
Total number of ${\vec \lambda}^0$ is equal to $m(m-1)/2$. 
Operating ${\wtilde R}^{0,0}(m; {\vec \lambda}^0)$ on the minimum weight states (\ref{4-6}), 
we define the states 
\bsub\label{4-15}
\beq
& &\ket{N;{\vec \lambda}^0,{\vec \rho},{\vec \sigma}^0}={\wtilde R}^{0,0}(m;{\vec \lambda}^0)
\ket{N;{\vec \rho},{\vec \sigma}^0}\ , 
\label{4-15a}\\
& &\ket{N;{\vec \lambda}^0,{\vec \rho},\rho^m,{\vec \sigma}^0}={\wtilde R}^{0,0}(m;{\vec \lambda}^0)
\ket{N;{\vec \rho},\rho^m,{\vec \sigma}^0}\ .
\label{4-15b}
\eeq
\esub
The states (\ref{4-15}) are the eigenstates of ${\wtilde R}_0(m;r)$ given in the relation (\ref{4-2}). 
After lengthy calculation, we can derive the commutation relation 
\beq\label{4-16}
\left[\ {\wtilde R}_0(m;r)\ , \ {\wtilde R}^{0,0}(m;r',k')\ \right]
=\left\{
\begin{array}{lr}
{\displaystyle \left(F_{r,r'}(m)-F_{r,k'}(m)\right){\wtilde R}^{0,0}(m;r',k')}\ , & \quad (4.16{\rm a}) \\
\qquad\qquad (r=1,2,\cdots , m-1) & \\
0\ . \qquad\ \  (r=m) & (4.16{\rm b})
\end{array}
\right.
\nonumber
\eeq
The relation (\ref{4-16}) is only applicable to the case with $n=2m+1$. 
With the use of the relation (\ref{4-16}), we can show that the states (\ref{4-15}) are 
the eigenstates of ${\wtilde R}_0(m;r)$, the eigenvalues of which we denote 
$\rho_0^r\ (r=1,2,\cdots , m-1, m$): 
\beq\label{4-17}
\rho_0^r=
\left\{
\begin{array}{lr}
{\displaystyle \sum_{r'=1}^{m-1}\left(\sum_{k'=r'+1}^{m}\left(
F_{r,r'}(m)-F_{r,k'}(m)\right)\lambda(m';k')\right)-\rho^r}\ , & 
\qquad\qquad\qquad (4.17{\rm a})\\
\qquad\qquad (r=1,2,\cdots , m-1) & \\
{\displaystyle -\rho^m}\ . \ \quad (r=m) & 
\qquad (4.17{\rm b})
\end{array}\right.
\nonumber
\eeq
In a form similar to that given in the relation (\ref{4-4}), we will use the notation ${\vec \rho}_0$: 
\setcounter{equation}{17}
\beq\label{4-18}
{\vec \rho}_0=\rho_0^1,\ \rho_0^2,\ \cdots , \ \rho_0^{m-1}\ . 
\eeq
Let us pick up two states specified by ${\vec \lambda}^0(1)$ and ${\vec \lambda}^0(2)$ with 
${\vec \lambda}^0(1)\neq {\vec \lambda}^0(2)$. 
If these two give us ${\vec \rho}\ ^0(1)$ and ${\vec \rho}\ ^0(2)$, respectively, 
with ${\vec \rho}\ ^0(1)\neq {\vec \rho}\ ^0(2)$, these two states are orthogonal to each other. 
But, if ${\vec \rho}\ ^0(1)={\vec \rho}\ ^0(2)$, they are not always orthogonal. 
However, they are linearly independent of each other, because the states (\ref{4-15}) 
are specified by ${\vec \lambda}^0$, the number of which is $m(m-1)/2$. 
It is equal to total of the quantum numbers characterizing the $su(m)$-algebra 
except the minimum weight states. 
In such cases, we must adopt some idea, for example, the Schmidt method for constructing the orthogonal set.

We must, further, investigate the role of the vectors and the spinors for constructing the linearly independent 
basis. 
These are defined in the relations (\ref{2-10})$\sim$(\ref{2-12}) and (\ref{2-14})$\sim$(\ref{2-17}). 
A simple idea for this task is to construct the raising operators by appropriate products of the vectors and the spinors, 
which are similar to the forms (\ref{4-5}) and (\ref{4-13}). 
With the help of this procedure, we can construct the linearly independent basis, which may be justified 
in the relations (\ref{2-23}) and (\ref{2-24}). 
However, for this task, we have a problem to be reconsidered. 
As was already mentioned, 
our present system contains two subalgebras, i.e., the $su(2)$- and the $su(m)$-algebra. 
In our form, the vectors and the spinors are introduced as the tensor operators in the $su(2)$-algebra. 
Until the present, we do not have contacted with the argument on the relation between the tensors and the $su(m)$-algebra. 
Therefore, not only from the side of the $su(2)$-algebra, but also from the side of the $su(m)$-algebra, 
it should be reconsidered. 
But, this consideration may be impossible to perform for the general case with arbitrary $m$. 
In next section, some concrete results for the cases with $n=2, 3, 4$ and 5 will be presented. 
On that occasion, we will have to contact with this point.

\setcounter{equation}{0}
\section{Some simple examples and discussion}

In {\bf 6} of (I), we showed the minimum weight states of the cases with $n=2,\ 3,\ 4$ and 5.
In this section, mainly, for the above cases, we will show the results obtained in this paper. 
First, will treat the $su(2)$-Lipkin model, i.e., $n=2$ and $m=1$. 
In this case, only one case with $r=1$ appears and the index $k$ is unnecessary. 
Three generators are given by 
\beq\label{5-1}
{\wtilde S}_+(1;1)={\wtilde S}^1\ (={\wtilde S}_+)\ , \quad
{\wtilde S}_-(1;1)={\wtilde S}_1\ (={\wtilde S}_-)\ , \quad
{\wtilde S}_0(1;1)=\frac{1}{2}{\wtilde S}_1^1\ (={\wtilde S}_0)\ . 
\eeq
The orthogonal set $\{\ket{ss_0}\}$ is given in the relation 
\beq\label{5-2}
\ket{ss_0}=\sqrt{\frac{(s-s_0)!}{(2s)!(s+s_0)!}}\left({\wtilde S}_+\right)^{s+s_0}\ket{s}\ . 
\eeq
Hereafter, we will omit the total fermion number $N$ and the notations used in {\bf 6} of (I) 
will be adopted, for example, ${\wtilde S}_{\pm, 0}$ in the relation (\ref{5-1}).

Next, we will discuss the $su(3)$-Lipkin model, i.e., $n=3$ and $m=1$. 
In this case, also the case with $r=1$ appears and $k$ is unnecessary. 
The eight generators are shown in the form 
\bsub\label{5-3}
\beq
& &{\wtilde S}_+(1;1)={\wtilde S}_1^2\ (={\wtilde S}_+)\ , \quad
{\wtilde S}_-(1;1)={\wtilde S}_2^1\ (={\wtilde S}_-)\ , \nonumber\\
& &
{\wtilde S}_0(1;1)=\frac{1}{2}\left({\wtilde S}_2^2-{\wtilde S}_1^1\right)\ (={\wtilde S}_0)\ , 
\qquad\ 
\label{5-3a}\\
& &{\wtilde R}^{\frac{1}{2},+\frac{1}{2}}(1;1)={\wtilde S}^2\ (={\wtilde R}^{\frac{1}{2},+\frac{1}{2}})\ , \quad
{\wtilde R}^{\frac{1}{2},-\frac{1}{2}}(1;1)={\wtilde S}^1\ (={\wtilde R}^{\frac{1}{2},-\frac{1}{2}})\ , \nonumber\\
& &{\wtilde R}_{\frac{1}{2},+\frac{1}{2}}(1;1)={\wtilde S}_2\ (={\wtilde R}_{\frac{1}{2},+\frac{1}{2}})\ , \quad
{\wtilde R}_{\frac{1}{2},-\frac{1}{2}}(1;1)={\wtilde S}_1\ (={\wtilde R}_{\frac{1}{2},-\frac{1}{2}})\ ,  
\label{5-3b}\\
& &{\wtilde {\cal P}}_0(1;1)=\frac{1}{2}\left({\wtilde S}_2^2+{\wtilde S}_1^1\right)\ (={\wtilde R}_0)\ , 
\label{5-3c}
\eeq
\esub
The minimum weight state $\ket{\rho,\sigma}$ obeys 
\beq\label{5-4}
{\wtilde S}_-\ket{\rho,\sigma}={\wtilde R}_{\frac{1}{2},\pm\frac{1}{2}}\ket{\rho,\sigma}=0\ , \quad
{\wtilde S}_0\ket{\rho,\sigma}=-\sigma\ket{\rho,\sigma}\ , \quad
{\wtilde R}_0\ket{\rho,\sigma}=-\rho\ket{\rho,\sigma}\ . 
\eeq
Therefore, we have the state with the eigenvalue $(\sigma,\sigma_0)$ in the form 
\beq\label{5-5}
\ket{\rho,\sigma\sigma_0}
=\sqrt{\frac{(\sigma-\sigma_0)!}{(2\sigma)!(\sigma+\sigma_0)!}}
\left({\wtilde S}_+\right)^{\sigma+\sigma_0}\ket{\rho,\sigma}\ . 
\eeq
With the use of the state (\ref{5-5}), we define the following state: 
\beq
& &\ket{j,ss_0;\rho\sigma}=\sum_{j_0\sigma_0}\langle jj_0\sigma\sigma_0\ket{ss_0}{\wtilde R}^{j,j_0}\ket{\rho,\sigma\sigma_0}\ , 
\label{5-6}\\
& &{\wtilde R}^{j,j_0}=\sqrt{\frac{(2j)!}{(j+j_0)!(j-j_0)!}}\left({\wtilde R}^{\frac{1}{2},+\frac{1}{2}}\right)^{j+j_0}
\left({\wtilde R}^{\frac{1}{2},-\frac{1}{2}}\right)^{j-j_0}\ . 
\label{5-7}
\eeq
Here, ${\wtilde R}^{j,j_0}$ denotes the tensor operator with rank $j$, i.e., $j=1/2,\ 1,\ 3/2,\cdots $ and 
$j_0=-j,\ -j+1,\cdots ,\ j-1,\ j$. 
It is easily verified that $\ket{j,ss_0;\rho\sigma}$ satisfies 
\bsub\label{5-8}
\beq
& &{\wtilde R}_0\ket{j,ss_0;\rho\sigma}=(3j-\rho)\ket{j,ss_0;\rho\sigma}\ , 
\label{5-8a}\\
& &{\wtilde {\mib S}}^2\ket{j,ss_0;\rho\sigma}=s(s+1)\ket{j,ss_0;\rho\sigma}\ , 
\label{5-8b}\\
& &{\wtilde S}_0\ket{j,ss_0;\rho\sigma}=s_0\ket{j,ss_0;\rho\sigma}\ , 
\label{5-8c}\\
& &{\wtilde \Gamma}_{su(3)}\ket{j,ss_0;\rho\sigma}=\left(\sigma(\sigma+1)+\frac{1}{3}\rho(\rho+3)\right)\ket{j,ss_0;\rho\sigma}\ . 
\label{5-8d}
\eeq
\esub
Therefore, we can learn that the set $\{\ket{j,ss_0;\rho\sigma}\}$ is orthogonal and, as is well known, this set is 
specified by five quantum numbers characterizing the $su(3)$-algebra.

Our third concern is related to the $su(4)$-Lipkin model, i.e., $n=4$ and $m=2$. 
For $r=1$ and 2, the $su(2)$-subalgebras can be expressed as 
\bsub\label{5-9}
\beq
& &{\wtilde S}_+(2;1)={\wtilde S}_2^3\ (={\wtilde S}_+(1)),\quad
{\wtilde S}_-(2;1)={\wtilde S}_3^2\ (={\wtilde S}_-(1)),\nonumber\\
& &
{\wtilde S}_0(2;1)=\frac{1}{2}\left({\wtilde S}_3^3-{\wtilde S}_2^2\right)\ (={\wtilde S}_0(1)),
\nonumber\\
& & 
\label{5-9a}\\
& &{\wtilde S}_+(2;2)={\wtilde S}^1\ (={\wtilde S}_+(2)),\quad
{\wtilde S}_-(2;2)={\wtilde S}_1\ (={\wtilde S}_-(2)),\nonumber\\
& &
{\wtilde S}_0(2;2)=\frac{1}{2}{\wtilde S}_1^1\ (={\wtilde S}_0(2)).
\label{5-9b}
\eeq
\esub 
The addition can be denoted as 
\beq\label{5-10}
{\wtilde S}_{\pm,0}={\wtilde S}_{\pm,0}(1)+{\wtilde S}_{\pm,0}(2)\ . 
\eeq
For the scalar and the vector operators, which appear in the case with $(r=1,\ k=2)$, are written down as 
\beq
& &{\wtilde R}^{0,0}(2;1,2)={\wtilde S}_1^3+{\wtilde S}^2\ (={\wtilde R}_+)\ ,\quad
{\wtilde R}_{0,0}(2;1,2)={\wtilde S}_3^1+{\wtilde S}_2\ (={\wtilde R}_-)\ , 
\label{5-11}\\
& &{\wtilde R}^{1,+1}(2;1,2)=-{\wtilde S}^3\ (={\wtilde R}^{1,+1})\ ,\quad
{\wtilde R}_{1,+1}(2;1,2)=-{\wtilde S}_3\ (={\wtilde R}_{1,+1})\ , 
\label{5-12}\\
& &{\wtilde R}^{1,0}(2;1,2)=\frac{1}{\sqrt{2}}\left({\wtilde S}_1^3-{\wtilde S}^2\right)\ (={\wtilde R}^{1,0})\ ,\nonumber\\
& &
{\wtilde R}_{1,0}(2;1,2)=\frac{1}{\sqrt{2}}\left({\wtilde S}_3^1-{\wtilde S}_2\right)\ (={\wtilde R}_{1,0})\ , 
\qquad\ 
\label{5-13}\\
& &{\wtilde R}^{1,-1}(2;1,2)={\wtilde S}_1^2\ (={\wtilde R}^{1,-1})\ ,\quad
{\wtilde R}_{1,-1}(2;1,2)={\wtilde S}_2^1\ (={\wtilde R}_{1,-1})\ . 
\label{5-14}
\eeq
The operator ${\wtilde {\cal P}}_0(2;1)$ can be expressed as 
\beq\label{5-15}
{\wtilde {\cal P}}_0(2;1)=\frac{1}{2}\left({\wtilde S}_3^3+{\wtilde S}_2^2-{\wtilde S}_1^1\right)\ (={\wtilde R}_0)\ . 
\eeq
The above are fifteen generators of the $su(4)$-algebra. 
In {\bf 4}, we mentioned that the scalar operators form the $su(m)$-algebra, in which the present case is given by 
the $su(2)$-algebra $({\wtilde R}_{\pm,0})$. 
Of course, we have 
\beq\label{5-16}
\left[\ {\rm any\ of}\ {\wtilde R}_{\pm,0}\ , \ {\rm any\ of}\ {\wtilde S}_{\pm,0}\ \right] =0\ .
\eeq
Then, as the hermitian operators commuting with the Casimir operator ${\wtilde \Gamma}_{su(4)}$, 
we can choose $({\wtilde {\mib S}}^2,{\wtilde S}_0)$ and $({\wtilde {\mib R}}^2,{\wtilde R}_0)$. 
By putting $\sigma_0=-\sigma$ in the 
state (I.6.13), we obtain the state $\ket{\rho,\sigma^1,\sigma^2,\sigma\sigma_0=-\sigma}$ and, then, 
the eigenstate for ${\wtilde {\mib R}}^2$, ${\wtilde R}_0$, ${\wtilde {\mib S}}^2$ and ${\wtilde S}_0$: 
\beq\label{5-17}
\ket{\rho\rho_0,\sigma^1,\sigma^2,\sigma\sigma_0}&=&
\sqrt{\frac{(\rho-\rho_0)!}{(2\rho)!(\rho+\rho_0)!}}\sqrt{\frac{(\sigma-\sigma_0)!}{(2\sigma)!(\sigma+\sigma_0)!}}\nonumber\\
& &\times \left({\wtilde R}_+\right)^{\rho+\rho_0}\left({\wtilde S}_+\right)^{\sigma+\sigma_0}\ket{\rho,\sigma^1,\sigma^2,\sigma\sigma_0=-\sigma}\ . 
\eeq
The present case is the simplest; $m=1$. 
Therefore, the state (\ref{5-17}) is easily obtained.

Now, we must discuss the vector operators (\ref{5-12})$\sim$(\ref{5-14}). 
As was already mentioned, there exist two $su(2)$-algebras satisfying the relation (\ref{5-16}). 
We started the present form in the $su(2)$-algebra $({\wtilde S}_{\pm,0})$. 
However, at the present stage, we have no reason why we give the treatment for $({\wtilde S}_{\pm,0})$ priority to that for 
$({\wtilde R}_{\pm,0})$. In other words, we must treat these two on an equal footing with each other. 
Although the vector operators (\ref{5-12})$\sim$({\ref{5-14}) are given for $({\wtilde S}_{\pm,0})$, they 
satisfy the relation 
\beq\label{5-18}
\left[\ {\wtilde R}_+\ , \ {\wtilde R}^{1,\nu}\ \right]=0\ , \quad
\left[\ {\wtilde R}_0\ , \ {\wtilde R}^{1,\nu}\ \right]=\nu{\wtilde R}^{1,\nu}\ . \quad
(\nu=\pm 1,\ 0)
\eeq
Then by calculating $[{\wtilde R}_-,{\wtilde R}^{1,\nu}]$ and $[{\wtilde R}_-,[{\wtilde R}_-,{\wtilde R}^{1,\nu}]]$, we can 
present the vector operators for $({\wtilde R}_{\pm,0})$: 
\beq
& &{\wtilde R}^{1,1;1,1}=-{\wtilde S}^3\ (={\wtilde R}^{1,1})\ , \nonumber\\
& &
{\wtilde R}^{1,1;1,0}=\frac{1}{\sqrt{2}}\left({\wtilde S}_2^3-{\wtilde S}^1\right)\ \left(=\frac{1}{\sqrt{2}}\left({\wtilde S}_+(1)-{\wtilde S}_+(2)\right)\right)\ , 
\nonumber\\
& &{\wtilde R}^{1,1;1,-1}={\wtilde S}_2^1\ (={\wtilde R}_{1,-1})\ , 
\label{5-19}\\
& &{\wtilde R}^{1,0;1,1}=\frac{1}{\sqrt{2}}\left({\wtilde S}_1^3-{\wtilde S}^2\right)\ (={\wtilde R}^{1,0})\ , \nonumber\\
& &
{\wtilde R}^{1,0;1,0}=-\frac{1}{\sqrt{2}}\left({\wtilde S}_3^3-{\wtilde S}_2^2-{\wtilde S}_1^1\right)\ \left(=-\left({\wtilde S}_0(1)-{\wtilde S}_0(2)\right)\right)\ , 
\nonumber\\
& &{\wtilde R}^{1,0;1,-1}=\frac{1}{\sqrt{2}}\left({\wtilde S}_3^1-{\wtilde S}_2\right)\ (={\wtilde R}_{1,0})\ . 
\label{5-20}\\
& &{\wtilde R}^{1,-1;1,1}={\wtilde S}_1^2\ (={\wtilde R}^{1,-1})\ , \nonumber\\
& &
{\wtilde R}^{1,-1;1,0}=-\frac{1}{\sqrt{2}}\left({\wtilde S}_3^2-{\wtilde S}_1\right)\ \left(=-\frac{1}{\sqrt{2}}\left({\wtilde S}_-(1)-{\wtilde S}_-(2)\right)\right)\ , 
\nonumber\\
& &{\wtilde R}^{1,-1;1,-1}=-{\wtilde S}_3\ (={\wtilde R}_{1,1})\ , 
\label{5-21}
\eeq
Here, ${\wtilde R}^{1,\mu;1,\nu}\ (\mu,\ \nu=\pm1, 0)$ denotes the tensor operator with $(1,\mu)$ for $({\wtilde S}_{\pm,0})$ and 
$(1,\nu)$ for $({\wtilde R}_{\pm,0})$. 
The operator ${\wtilde R}^{1,\mu;1,\nu}$ satisfies 
\beq\label{5-22}
\left({\wtilde R}^{1,\mu;1,\nu}\right)^*=(-)^{\mu+\nu}{\wtilde R}_{1,-\mu;1,-\nu}\ . 
\eeq
Therefore, we have 
\beq\label{5-23}
& &
\sum_{\mu\nu}(-)^{1-\mu}(-)^{1-\nu}{\wtilde R}^{1,\mu;1,\nu}{\wtilde R}^{1,-\mu;1,-\nu}
=\sum_{\mu\nu}{\wtilde R}^{1,\mu;1,\nu}{\wtilde R}_{1,\mu;1,\nu}
=\sum_{\mu\nu}{\wtilde R}_{1,\mu;1,\nu}{\wtilde R}^{1,\mu;1,\nu}
\nonumber\\
&=&2{\wtilde \Gamma}_{su(4)}-{\wtilde {\mib S}}^2-{\wtilde {\mib R}}^2\ . 
\eeq
Here, we used the relations (\ref{5-19})$\sim$(\ref{5-21}).

In order to obtain the linearly independent basis for the orthogonal set, for the $su(4)$-algebra, 
the basis should be expressed in terms of nine parameters 
(if possible, quantum numbers) including $(\rho,\sigma^1,\sigma^2)$. 
In the state $\ket{\rho\rho_0,\sigma^1,\sigma^2,\sigma\sigma_0}$ given in the relation (\ref{5-17}), 
six quantum numbers appear. 
Therefore, the linearly independent basis may be obtained by operating certain operator expressed in terms of three parameters on the 
state (\ref{5-17}). 
For searching this operator, it may be important to treat $({\wtilde R}_{\pm,0})$ on an equal footing with $({\wtilde S}_{\pm,0})$. 
First, we note the relation 
\bsub\label{5-24}
\beq
& &\left[\ {\wtilde S}_+\ , \ \left(-{\wtilde S}^3\right)^l\ \right]=\left[\ {\wtilde R}_+\ , \ \left(-{\wtilde S}^3\right)^l\ \right]=0\ , 
\label{5-24a}\\
& &\left[\ {\wtilde S}_0\ , \ \left(-{\wtilde S}^3\right)^l\ \right]=\left[\ {\wtilde R}_0\ , \ \left(-{\wtilde S}^3\right)^l\ \right]
=l\left(-{\wtilde S}^3\right)^l\ . \quad
(l=0,1,2, \cdots)
\label{5-24b}
\eeq
\esub
The relation (\ref{5-24}) suggests us that we can construct the tensor operator with rank $l$ which is common to 
$({\wtilde S}_{\pm,0})$ and $({\wtilde R}_{\pm,0})$ and if we denote it as 
${\wtilde R}^{l,l_0;l,\lambda_0}$, it can be given in the form 
\beq\label{5-25}
{\wtilde R}^{l,l_0;l,\lambda_0}=\sqrt{\frac{(l+l_0)!(l-l_0)!}{(2l)!}}\sqrt{\frac{(l+\lambda_0)!(l-\lambda_0)!}{(2l)!}}
\left({\poon R}_-\right)^{l-\lambda_0}\left({\poon S}_-\right)^{l-l_0}\left(-{\wtilde S^3}\right)^l\ . \qquad
\eeq
Here, we used the notations for ${\wtilde O}$ and ${\wtilde A}$ in the form 
\beq\label{5-26}
\left({\poon O}\right)^n{\wtilde A}
=\underbrace{
\left[{\wtilde O}, \cdots ,\left[{\wtilde O},\left[{\wtilde O}\right.\right.\right.}_{n},{\wtilde A}\Bigl]\Bigl]\cdots \Bigl]\ . 
\eeq
Then, operating ${\wtilde R}^{l,l_0;l,\lambda_0}$ on the state $\ket{\rho\rho_0,\sigma^1,\sigma^2,\sigma\sigma_0}$, we have 
\beq\label{5-27}
\ket{l\lambda_0,ll_0,\rho\rho_0,\sigma^1,\sigma^2,\sigma\sigma_0}
={\wtilde R}^{l,l_0;l,\lambda_0}\ket{\rho\rho_0,\sigma^1,\sigma^2,\sigma\sigma_0}\ . 
\eeq
Certainly, ${\wtilde R}^{l,l_0;l,\lambda_0}$ is expressed in terms of three parameters. 
With the use of the Clebsch-Gordan coefficients, we obtain the eigenstate of ${\wtilde {\mib S}}^2$, ${\wtilde S}_0$, 
${\wtilde {\mib R}}^2$, ${\wtilde R}_0$ and ${\wtilde \Gamma}_{su(4)}$: 
\beq\label{5-28}
& &\ket{l,rr_0,ss_0,\sigma;\rho\sigma^1\sigma^2}
=\sum_{\lambda_0\rho_0}\sum_{l_0\sigma_0}\langle l\lambda_0\rho\rho_0\ket{rr_0}
\langle ll_0\sigma\sigma_0\ket{ss_0}\ket{l\lambda_0,ll_0,\rho\rho_0,\sigma^1,\sigma^2,\sigma\sigma_0}\ , 
\nonumber\\
& &l=|r-\rho|,\cdots ,\ (r+\rho)\quad {\rm and}\quad |s-\sigma|,\cdots ,\ (s+\sigma)\ . 
\eeq
The eigenvalues are given by $s(s+1)$, $s_0$, $r(r+1)$, $r_0$ and $\sigma^1(\sigma^1+1)+\sigma^2(\sigma^2+1)+\rho(\rho+4)/2$, respectively. 
Except $l$, the eight parameters are the quantum numbers. 
Of course, $\sigma=|\sigma^1-\sigma^2|,\cdots,\ (\sigma^1+\sigma^2)$. 
Therefore, the set composed of the states (\ref{5-28}) may be the linearly independent basis for the $su(4)$-Lipkin model.

Finally, we will discuss the case with $n=5$, i.e., $m=2$. 
This case contains also two $su(2)$-algebra: 
\bsub\label{5-29}
\beq
& &{\wtilde S}_+(2;1)={\wtilde S}_3^4\ (={\wtilde S}_+(1)) ,\ \ 
{\wtilde S}_-(2;1)={\wtilde S}_4^3\ (={\wtilde S}_-(1)) ,\nonumber\\
& &
{\wtilde S}_0(2;1)=\frac{1}{2}\left({\wtilde S}_4^4-{\wtilde S}_3^3\right)\ (={\wtilde S}_0(1)) ,\nonumber\\
& &
\label{5-29a}\\
& &{\wtilde S}_+(2;2)={\wtilde S}_1^2\ (={\wtilde S}_+(2)),\ \ 
{\wtilde S}_-(2;2)={\wtilde S}_2^1\ (={\wtilde S}_-(2)) ,\nonumber\\
& & 
{\wtilde S}_0(2;2)=\frac{1}{2}\left({\wtilde S}_2^2-{\wtilde S}_1^1\right)\ (={\wtilde S}_0(2)) . \nonumber\\
& &\label{5-29b}
\eeq
\esub
Then,the addition of the two is expressed as 
\beq\label{5-30}
{\wtilde S}_{\pm,0}={\wtilde S}_{\pm,0}(1)+{\wtilde S}_{\pm,0}(2)\ .
\eeq
The scalar operators are as follows: 
\beq\label{5-31}
{\wtilde R}^{0,0}(2;1,2)={\wtilde S}_2^4+{\wtilde S}_1^3\ (={\wtilde R}_+)\ , \quad 
{\wtilde R}_{0,0}(2;1,2)={\wtilde S}_4^2+{\wtilde S}_3^1\ (={\wtilde R}_-)\ .
\eeq
In the present case, two spinors appear: 
\bsub\label{5-32}
\beq
& &{\wtilde R}^{\frac{1}{2},+\frac{1}{2}}(2;1)={\wtilde S}^4\ (={\wtilde R}^{\frac{1}{2},+\frac{1}{2}}(1))\ , \quad
{\wtilde R}^{\frac{1}{2},-\frac{1}{2}}(2;1)={\wtilde S}^3\ (={\wtilde R}^{\frac{1}{2},-\frac{1}{2}}(1))\ , \nonumber\\
& &{\wtilde R}_{\frac{1}{2},+\frac{1}{2}}(2;1)={\wtilde S}_4\ (={\wtilde R}_{\frac{1}{2},+\frac{1}{2}}(1))\ , \quad
{\wtilde R}_{\frac{1}{2},-\frac{1}{2}}(2;1)={\wtilde S}_3\ (={\wtilde R}_{\frac{1}{2},-\frac{1}{2}}(1))\ ,
\label{5-32a}\\
& &{\wtilde R}^{\frac{1}{2},+\frac{1}{2}}(2;2)={\wtilde S}^2\ (={\wtilde R}^{\frac{1}{2},+\frac{1}{2}}(2))\ , \quad
{\wtilde R}^{\frac{1}{2},-\frac{1}{2}}(2;2)={\wtilde S}^1\ (={\wtilde R}^{\frac{1}{2},-\frac{1}{2}}(2))\ , \nonumber\\
& &{\wtilde R}_{\frac{1}{2},+\frac{1}{2}}(2;2)={\wtilde S}_2\ (={\wtilde R}_{\frac{1}{2},+\frac{1}{2}}(2))\ , \quad
{\wtilde R}_{\frac{1}{2},-\frac{1}{2}}(2;2)={\wtilde S}_1\ (={\wtilde R}_{\frac{1}{2},-\frac{1}{2}}(2))\ .
\label{5-32b}
\eeq
\esub
The vector operators are given as 
\beq
& &{\wtilde R}^{1,+1}(2;1,2)=-{\wtilde S}_1^4\ (={\wtilde R}^{1,+1})\ , \quad
{\wtilde R}_{1,+1}(2;1,2)=-{\wtilde S}_4^1 \ (={\wtilde R}_{1,+1})\ , 
\label{5-33}\\ 
& &{\wtilde R}^{1,+0}(2;1,2)=\frac{1}{\sqrt{2}}\left({\wtilde S}_2^4-{\wtilde S}_1^3\right)\ , \quad
{\wtilde R}_{1,0}(2;1,2)=\frac{1}{\sqrt{2}}\left({\wtilde S}_4^2-{\wtilde S}_3^1\right)\ , 
\label{5-34}\\
& &{\wtilde R}^{1,-1}(2;1,2)={\wtilde S}_2^3\ (={\wtilde R}^{1,-1})\ , \quad
{\wtilde R}_{1,-1}(2;1,2)={\wtilde S}_3^2 \ (={\wtilde R}_{1,-1})\ , 
\label{5-35}
\eeq
The operators ${\wtilde {\cal P}}_0(2;1)$ and ${\wtilde {\cal P}}_0(2;2)$ can be expressed as 
\beq\label{5-36}
{\wtilde {\cal P}}_0(2;1)=\frac{1}{2}\left({\wtilde S}_4^4+{\wtilde S}_3^3\right)\ , \qquad
{\wtilde {\cal P}}_0(2;2)=\frac{1}{2}\left({\wtilde S}_2^2+{\wtilde S}_1^1\right)\ ,
\eeq
i.e.,
\bsub\label{5-37}
\beq
& &{\wtilde R}_0(1)={\wtilde {\cal P}}_0(2;1)-{\wtilde {\cal P}}_0(2;2)=
\frac{1}{2}\left({\wtilde S}_4^4+{\wtilde S}_3^3-{\wtilde S}_2^2-{\wtilde S}_1^1\right)\ (={\wtilde R}_0)\ , 
\label{5-37a}\\
& &{\wtilde R}_0(2)={\wtilde {\cal P}}_0(2;1)+{\wtilde {\cal P}}_0(2;2)=
\frac{1}{2}\left({\wtilde S}_4^4+{\wtilde S}_3^3+{\wtilde S}_2^2+{\wtilde S}_1^1\right)\ .
\label{5-37b}
\eeq
\esub
With the use of the above relations, we will show the results for the case with $n=5$ simply in comparative argument with the case with $n=4$.

In (I), the minimum weight state in the present case is given by $\ket{\rho^1,\rho^2,\sigma^1,\sigma^2}$, where 
$\rho^1$, $\rho^2$, $\sigma^1$ and $\sigma^2$ denote the eigenvalues of 
${\wtilde R}_0(1)\ (={\wtilde R}_0)$, ${\wtilde R}_2(2)$, ${\wtilde S}_0(1)$ and ${\wtilde S}_0(2)$, respectively. 
The set $({\wtilde R}_{\pm,0})$ also forms the $su(2)$-algebra obeying the same relation as that given in the relation (\ref{5-16}) and there 
exists also one set of the vector for $({\wtilde S}_{\pm,0})$. 
Therefore, the treatment for these two may be similar to the case with $n=4$. 
The difference of these two can be found in the appearance of the hermitian operator ${\wtilde R}_0(2)$ 
and two sets of the spinors. 
Under the same idea as that in the case with $n=4$, we introduce ${\wtilde R}^{\frac{1}{2},\mu;\frac{1}{2},\nu}$: 
\beq\label{5-38}
& &{\wtilde R}^{\frac{1}{2},+\frac{1}{2};\frac{1}{2},+\frac{1}{2}}={\wtilde S}^4,\quad
{\wtilde R}^{\frac{1}{2},-\frac{1}{2};\frac{1}{2},+\frac{1}{2}}={\wtilde S}^3,\quad
{\wtilde R}^{\frac{1}{2},+\frac{1}{2};\frac{1}{2},-\frac{1}{2}}={\wtilde S}^2,\quad
{\wtilde R}^{\frac{1}{2},-\frac{1}{2};\frac{1}{2},-\frac{1}{2}}={\wtilde S}^1.\nonumber\\
& &
\eeq 
Then, we can define the following scalar operator for $({\wtilde S}_{\pm,0})$ and $({\wtilde R}_{\pm,0})$: 
\beq\label{5-39}
{\wtilde R}^{0,0;0,0}=\frac{1}{2}\sum_{\mu\nu}(-)^{\frac{1}{2}-\mu}(-)^{\frac{1}{2}-\nu}
{\wtilde R}^{\frac{1}{2},\mu;\frac{1}{2},\nu}{\wtilde R}^{\frac{1}{2},-\mu;\frac{1}{2},-\nu}
={\wtilde S}^4{\wtilde S}^1-{\wtilde S}^3{\wtilde S}^2\ .
\eeq
The above should be compared with the relation (\ref{5-23}). 
Further, we have the operator ${\wtilde R}^{j,j_0;j,\kappa_0}$:
\beq\label{5-40}
{\wtilde R}^{j,j_0;j,\kappa_0}
&=&\sqrt{\frac{(j+j_0)!(j-j_0)!}{(2j)!}}\cdot\sqrt{\frac{(j+\kappa_0)!(j-\kappa_0)!}{(2j)!}}
\left({\poon R}_-\right)^{j-\kappa_0}\left({\poon S}_-\right)^{j-j_0}\left({\wtilde S}^4\right)^{2j}\ . 
\nonumber\\
& &\qquad\qquad\qquad\qquad\qquad\qquad\qquad\qquad\qquad
\left(j=0, \ \frac{1}{2},\  1, \ \frac{3}{2},\cdots \right)
\eeq
The operators (\ref{5-39}) and (\ref{5-40}) satisfy 
\beq\label{5-41}
\left[\ {\wtilde R}_0(2)\ , \ \left({\wtilde R}^{0,0;0,0}\right)^{n_0}{\wtilde R}^{j,j_0;j,\kappa_0}\ \right]=
5(n_0+j)\left({\wtilde R}^{0,0;0,0}\right)^{n_0}{\wtilde R}^{j,j_0;j,\kappa_0}\ . 
\eeq
In this connection, we have 
\beq\label{5-42}
& &\left[\ {\wtilde R}_0(2)\ , \ {\wtilde R}^{l,l_0;l,\lambda_0}\ \right]=
\left[\ {\wtilde R}_0(2)\ , \ {\wtilde R}_+\ \right]=\left[\ {\wtilde R}_0(2)\ , \ {\wtilde S}_+(1)\ \right]=
\left[\ {\wtilde R}_0(2)\ , \ {\wtilde S}_+(2)\ \right]=0 \ . \nonumber\\ 
& &
\eeq
Of course, ${\wtilde R}^{l,l_0;l,\lambda_0}$ is obtained by replacing $(-{\wtilde S}^3)$ with ${\wtilde S}^4$ in the relation (\ref{5-25}).

The above operators give us the following state: 
\beq\label{5-43}
& &\ket{ll_0\lambda_0,n_0jj_0\kappa_0,\rho^1\rho_0^1,\rho^2,\sigma^1\sigma_0^1,\sigma^2\sigma_0^2}\nonumber\\
&=&\!\!\! \left({\wtilde R}^{l,l_0;l,\lambda_0}\right)\left[\left({\wtilde R}^{0,0;0,0}\right)^{n_0}{\wtilde R}^{j,j_0;j,\kappa_0}\right]
\nonumber\\
& &\times
\left({\wtilde R}_+\right)^{\rho^1+\rho_0^1}\left({\wtilde S}_+(1)\right)^{\sigma^1+\sigma_0^1}
\left({\wtilde S}_+(2)\right)^{\sigma^2+\sigma_0^2}
\ket{\rho^1,\rho^2,\sigma^1,\sigma^2}\ . 
\eeq
We can see that the state (\ref{5-43}) is expressed in terms of fourteen parameters including $\rho^1$, $\rho^2$, $\sigma^1$ and $\sigma^2$. 
By using the angular momentum coupling rule, the state (\ref{5-43}) can be rewritten to the form with 
the eigenvalues of ${\wtilde {\mib S}}^2$, ${\wtilde S}_0$, ${\wtilde {\mib R}}^2$ and ${\wtilde R}_0$, which is 
similar to the state (\ref{5-28}). 
Of course, the states (\ref{5-43}) form the linearly independent basis for the $su(5)$-Lipkin model. 
The relations (\ref{5-41}) and (\ref{5-42}) teach us that the state (\ref{5-43}) is the eigenstate of ${\wtilde R}_0(2)$ with the 
eigenvalue $5(n_0+j)-\rho^2$. 
The above is an outline of our idea for the $su(5)$-Lipkin model.

Through the above arguments, we can learn some characteristic features of the Lipkin model. 
Classification of the model into two cases with $n=2m$ and $2m+1$ may be essential for understanding the features demonstrated by the model. 
Number of the $su(2)$-subalgebras is equal to $m$ for both cases and, further, they contain the $su(m)$-subalgebra. 
In the case with $n=2m+1$, there exists the degree of freedom related with ${\wtilde R}_0(m)$ and its eigenvalues are 
influenced by the spinors which do not exist in the case with $n=2m$.

\setcounter{equation}{0}
\section{Concluding remarks}

In the present two papers, we discussed formal aspects of the $su(n)$-Lipkin model in rather general form. 
In (I), we are mainly concerned with the determination of the minimum weight states. 
In (II), after re-forming the model in the frame of spherical tensor representation, we re-formulated the model by classifying it into two cases: 
$n=2m$ and $2m+1$. 
Especially, in {\S 5}, we gave some concrete results for the cases with $n=2,3,4$ and 5, which, in (I), we discussed only the minimum weight states. 
As a concluding remark, we will show connection between the $su(2m)$- and the $su(2m+1)$-Lipkin model directly.

%
%%%%%%%%%%%%%%%%%%%%%%%%%%%%%%%%%%%%%%%%%%%%%%%%%%%%%%%%%%%%%%%%%%%%%%
\begin{figure}[b]
\begin{center}
\includegraphics[height=5.5cm]{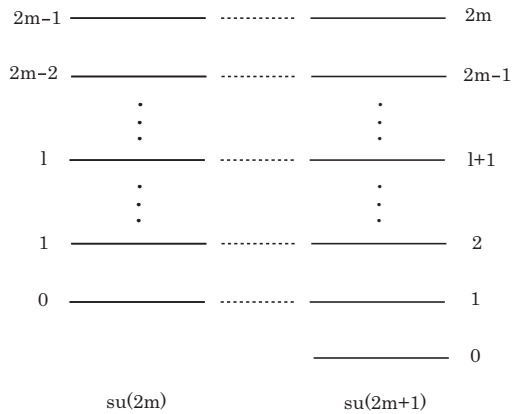}
\caption{The single-particle level schemes are schematically depicted in the case of the 
$su(2m)$-Lipkin model (left) 
and the $su(2m+1)$-Lipkin model (right), respectively. 
}
\label{fig:fig1}
\end{center}
\end{figure}
%%%%%%%%%%%%%%%%%%%%%%%%%%%%%%%%%%%%%%%%%%%%%%%%%%%%%%%%%%%%%%%%%%%%%%%%
%

Concerning the single-particle level schemes, we regard these two as the models illustrated in Fig.\ref{fig:fig1}. 
Let the single-particle level specified by the integer $l$ ($l=0, 1,2\cdots , 2m-1$) in the 
$su(2m)$-Lipkin model be shifted to the integer $(l+1)$ in the 
$su(2m+1)$-Lipkin model. 
Therefore, the zero-th level cannot get any counterpart in the $su(2m)$-Lipkin model. 
As a possible candidate, the following correspondence may be permitted: 
\beq\label{6-1}
{\wtilde S}^p \longrightarrow {\wtilde S}_1^{p+1}\ , \quad
{\wtilde S}_p \longrightarrow {\wtilde S}_{p+1}^1\ , \quad
{\wtilde S}_q^p \longrightarrow {\wtilde S}_{q+1}^{p+1}-\delta_{pq}{\wtilde S}_1^1\ . 
\quad
(p,\ q=1,\ 2,\cdots ,\ 2m-1)
\eeq
The set $\{\ {\wtilde S}_1^{p+1},\ {\wtilde S}_{p+1}^1,\ {\wtilde S}_{q+1}^{p+1}\ \}$ forms
the $su(2m)$-algebra and it becomes the $u(2m)$-algebra, if $\sum_{p=1}^{2m}{\wtilde S}_p^p$ is included. 
Further, if the operators ${\wtilde S}^p$ and ${\wtilde S}_p$ ($p=1,2,\cdots , 2m)$ are included, it becomes the $su(2m+1)$-algebra. 
In fact, total number of the generators is given by 
\beq\label{6-2}
[ (2m)^2-1 ]+1+2\cdot(2m)=(2m+1)^2-1\ . 
\eeq
Under the correspondence (\ref{6-1}), ${\wtilde \Gamma}_{su(2m)}$ given in the relation (\ref{3-1}) leads us to 
\beq\label{6-3}
{\wtilde \Gamma}_{su(2m)}
\longrightarrow 
\frac{1}{2}\left[
\sum_{p=2}^{2m}\sum_{q=1}^{p-1}\left[\ {\wtilde S}_q^p\ , \ {\wtilde S}_p^q\ \right]_+
+\sum_{p=1}^{2m}\left({\wtilde S}_p^p\right)^2-\frac{1}{2m}\left(\sum_{p=1}^{2m}{\wtilde S}_p^p\right)^{\!\!\!2}\right]\ .
\eeq
Adding the operators ${\wtilde S}^p$, ${\wtilde S}_p$ and $\sum_{p=1}^{2m}{\wtilde S}_p^p$ to the form 
(\ref{6-3}) in terms of\break
$\sum_{p=1}^{2m}[\ {\wtilde S}^p \ , \ {\wtilde S}_p\ ]_+/2$ and 
$(1/(2m(2m+1))(\sum_{p=1}^{2m}{\wtilde S}_p^p)^2/2$, we obtain ${\wtilde \Gamma}_{su(2m+1)}$.

As the $su(m)$-subalgebra, we treated the simplest case; $m=2$. 
In near future, we will report some results in the cases with $m>2$. 
It may be interesting to investigate how the cases with $m=3$ and 4 related with the Lipkin models with $n=3$ and 4, 
which were discussed in this paper.

\section*{Acknowledgment}

%One of the authors (M.Y.) wishes to express his appreciation to Mrs. Y. Miyamoto 
%for her hearty encouragement.  

Two of the authors (Y.T. and M.Y.) would like to express their thanks to 
Professor J. da Provid\^encia and Professor C. Provid\^encia, two of co-authors of this paper, 
for their warm hospitality during their visit to Coimbra in spring of 2015. 
The author, M.Y., would like to express his sincere thanks to Mrs K. Yoda-Ono for her cordial encouragement. 
The authors, Y.T., is partially supported by the Grants-in-Aid of the Scientific Research 
(No.26400277) from the Ministry of Education, Culture, Sports, Science and 
Technology in Japan.

\appendix
\section{The $su(m)$-algebra in the $su(n)$-Lipkin model}

The aim of Appendix A is to show that, as a subalgebra, the $su(n)$-Lipkin model contains the $su(m)$-algebra. 
Discussion starts in the case with $n=2m$. 
We define the following set of the operators specified by $p$ and $q$ $(p,\ q=1, \ 2,\cdots ,\ m-1)$:
\bsub\label{a1}
\beq
& &{\wtilde R}^p(m)={\wtilde S}_1^{2m-2p+1}+{\wtilde S}^{2m-2p}\ , \qquad
{\wtilde R}_p(m)={\wtilde S}_{2m-2p+1}^1+{\wtilde S}_{2m-2p}\ , 
\label{a1a}\\
& &{\wtilde R}_q^p(m)={\wtilde S}_{2m-2q+1}^{2m-2p+1}+{\wtilde S}_{2m-2q}^{2m-2p}-\delta_{pq}{\wtilde S}_1^1\ . 
\label{a1b}
\eeq
\esub
The above set obeys the commutation relation 
\bsub\label{a2}
\beq
& &[\ {\wtilde R}^p(m)\ , \ {\wtilde R}_q(m)\ ]={\wtilde R}_q^p(m)\ , 
\label{a2a}\\
& &[\ {\wtilde R}_q^p(m)\ , \ {\wtilde R}^r(m)\ ]=\delta_{qr}{\wtilde R}^p(m)+\delta_{pq}{\wtilde R}^r(m)\ , 
\label{a2b}\\
& &[\ {\wtilde R}_q^p(m)\ , \ {\wtilde R}_r^s(m)\ ]=\delta_{qs}{\wtilde R}_r^p(m)-\delta_{pr}{\wtilde R}_q^s(m)\ . 
\label{a2c}
\eeq
\esub
If ${\wtilde R}$ and $m$ in the relation (\ref{a2}) read ${\wtilde S}$ and $n$ in the relation (I.2.3), respectively, 
we can see that the set (\ref{a1}) obeys the $su(m)$-algebra. 
Of course, total number of the generators is equal to 
$(m-1)^2+2(m-1)=m^2-1$.

In the case with $p<q$, let $p$ and $q$ in the relation (\ref{a1a}) read $r$ and $k$, 
respectively, $(r=1,\ 2,\cdots, \ m-1,\ k=r+1,\ r+2,\cdots ,\ m-1)$. 
Then, ${\wtilde R}_q^p(m)$ becomes ${\wtilde R}^{0,0}(m;r,k)$ given in the 
relation (\ref{2-9}) for the case with $k=r+1,\ r+2,\cdots ,\ m-1$. 
Further, we can regard ${\wtilde R}^p(m)$ as ${\wtilde R}^{0,0}(m;r,m)$. 
In the case with $p>q$, if $p$ and $q$ are exchanged each other, we have the 
hermitian conjugate of ${\wtilde R}^{0,0}(m;r,k)$, i.e., 
${\wtilde R}_{0,0}(m;r,k)$. 
In the case with $p=q$, $p$ reads $r$ and, then, ${\wtilde R}_p^p(m)$ reduces to $2{\wtilde {\cal P}}_0(m;r)$ given in the relation (\ref{3-22}). 
Of course, ${\wtilde {\cal P}}_0(m;r)$ is essentially equivalent to ${\wtilde R}_0(m;r)$. 
Thus, in the case with $n=2m$, we have the $su(m)$-subalgebra in the $su(n)$-Lipkin model.

Next, we discuss the case with $n=2m+1$. 
First, we define the following set of the operators specified by $p$ and $(p,\ q=1,\ 2,\cdots ,\ m)$:
\beq\label{a3}
{\wtilde R}_q^p(m)={\wtilde S}_{2m-2q+2}^{2m-2p+2}+{\wtilde S}_{2m-2q+1}^{2m-2p+1}\ . 
\eeq
We have the commutation relation 
\beq\label{a4}
[\ {\wtilde R}_q^p(m)\ , \ {\wtilde R}_r^s(m)\ ]=\delta_{qs}{\wtilde R}_r^p(m)-\delta_{pr}{\wtilde R}_q^s(m)\ . 
\eeq
The relation (\ref{a4}) tells us that the set $\{{\wtilde R}_q^p(m)\}$ obeys well-known $u(m)$-algebra. 
In order to get the $su(m)$-algebra, we define the following operator:
\beq\label{a5}
{\ovl R}_q^p(m)={\wtilde R}_q^p(m)-\delta_{pq}\frac{1}{m}\sum_{r=1}^{m}{\wtilde R}_r^r(m)\ .
\eeq
The set (\ref{a5}) satisfies 
\bsub\label{a6}\beq
& &[\ {\ovl R}_q^p(m)\ , \ {\ovl R}_r^s(m)\ ]=\delta_{qs}{\ovl R}_r^p(m)-\delta_{pr}{\ovl R}_q^s(m)\ , 
\label{a6a}\\
& &\sum_{p=1}^{m}{\ovl R}_p^p(m)=0\ . 
\label{a6b}
\eeq
\esub
The relation (\ref{a6b}) tells us that, in the type ${\ovl R}_p^p(m)$, we can choose 
$(m-1)$ independent operators. 
Therefore, the set $\{{\ovl R}_q^p(m)\}$ obeys the $su(m)$-algebra. 
Of course, total number of the generators is equal to $m^2-1$.

In the case with $p<q$, if $p$ and $q$ read $r$ and $k$, respectively, ${\ovl R}_q^p(m)\ (={\wtilde R}_q^p(m))$ becomes ${\wtilde R}^{0,0}(m;r,k)$. 
In the case with $p>q$, if $p$ and $q$ are 
exchanged each other, we obtain the hermitian conjugate of ${\wtilde R}^{0,0}(m;r,k)$, i.e., 
${\wtilde R}_{0,0}(m;r,k)$. 
In the case with $p=q$, $p$ reads $r$ and, then, ${\ovl R}_p^p(m)$ reduces to 
$2({\wtilde P}_0(m;r)-(1/m)\cdot\sum_{r'=1}^{m}{\wtilde P}_0(m;r'))$. 
With the use of $C_{r,r'}(m)$ given in the relation (\ref{3-10}), the relation (\ref{a5}) gives us 
\beq\label{a7}
\sum_{r'=1}^{m}C_{r,r'}(m){\ovl R}_{r'}^{r'}(m)=
2\left(1-\delta_{r,m}\right){\wtilde Q}_0(m;r)\ . 
\eeq
Therefore, we have 
\beq\label{a8}
{\wtilde Q}_0(m;r)=\frac{1}{2}\sum_{r'=1}^{m}C_{r,r'}(m){\ovl R}_{r'}^{r'}(m) \qquad {\rm for}\quad r=1,\ 2, \cdots ,\ m-1\ .
\eeq
For ${\wtilde Q}_0(m;m)=(\sqrt{m})^{-1}\sum_{r=1}^{m}{\wtilde R}_r^r(m)$, 
the following relation is derived: 
\beq\label{a9}
[\ {\wtilde R}_q^p(m)\ , \ {\wtilde Q}_0(m;m)\ ]=0\qquad {\rm for\ any}\ p\ {\rm and}\ q\ .
\eeq
The operator ${\wtilde Q}_0(m;r)$ is essentially equivalent to ${\wtilde R}_0(m;r)$. 
The operator ${\wtilde Q}_0(m;m)$ plays a role similar to total fermion number ${\wtilde N}(n)$ in the 
$su(n)$-Lipkin model. 
Thus, we know that the case with $n=2m+1$ contains the $su(m)$-algebra in the $su(n)$-Lipkin model.

\end{document}